\documentclass[]{pasj01}
\usepackage{rotating}
\draft

\begin{document} 
\Received{}
\Accepted{}

\title{ALMA TWENTY-SIX ARCMIN$^2$ SURVEY OF GOODS-S AT ONE-MILLIMETER (ASAGAO): MILLIMETER PROPERTIES OF STELLAR MASS SELECTED GALAXIES}

\author{Yuki \textsc{Yamaguchi}\altaffilmark{1}%
}
\altaffiltext{1}{Institute of Astronomy, Graduate School of Science, The University of Tokyo, 2-21-1 Osawa, Mitaka, Tokyo 181-0015, Japan}
\email{yyamaguchi@ioa.s.u-tokyo.ac.jp}

\author{Kotaro \textsc{Kohno},\altaffilmark{1,2}}
\altaffiltext{2}{Research Center for the Early Universe, Graduate School of Science, The University of Tokyo, 7-3-1, Hongo, Bunkyo, Tokyo 113-0033, Japan}

\author{Bunyo \textsc{Hatsukade}\altaffilmark{1}}

\author{Tao \textsc{Wang}\altaffilmark{1,3}}
\altaffiltext{3}{National Astronomical Observatory of Japan, 2-21-1 Osawa, Mitaka, Tokyo 181-8588, Japan}

\author{Yuki \textsc{Yoshimura}\altaffilmark{1}}

\author{Yiping \textsc{Ao}\altaffilmark{4,5}}
\altaffiltext{4}{Purple Mountain Observatory and Key Laboratory for Radio Astronomy, Chinese Academy of Sciences, 8 Yuanhua Road, Nanjing 210034, China}
\altaffiltext{5}{School of Astronomy and Space Science, University of Science and Technology of China, Hefei, Anhui, China}

\author{James S.~\textsc{Dunlop}\altaffilmark{6}}
\altaffiltext{6}{Institute for Astronomy, University of Edinburgh, Royal Observatory, Edinburgh EH9 3HJ, UK}

\author{Eiichi \textsc{Egami}\altaffilmark{7}}
\altaffiltext{7}{Steward Observatory, University of Arizona, 933 North Cherry Avenue, Tucson, AZ 85721, USA}

\author{Daniel \textsc{Espada}\altaffilmark{3,8,9}}
\altaffiltext{8}{Department of Astronomical Science, SOKENDAI (The Graduate University of Advanced Studies), 2-21-1 Osawa, Mitaka, Tokyo 181-8588, Japan}
\altaffiltext{9}{SKA Organization, Lower Withington, Macclesfield, Cheshire SK11 9DL, UK}

\author{Seiji \textsc{Fujimoto}\altaffilmark{10,3,11,12,13}}
\altaffiltext{10}{Institute for Cosmic Ray Research, The University of Tokyo, Kashiwa, Chiba 277-8582, Japan}
\altaffiltext{11}{Research Institute for Science and Engineering, Waseda University, 3-4-1 Okubo, Shinjuku, Tokyo 169-8555, Japan}
\altaffiltext{12}{Cosmic DAWN Center, Copenhagen, Denmark}
\altaffiltext{13}{Niels Bohr Institute, University of Copenhagen, Lyngbyvej 2, DK-2100, Copenhagen, Denmark}

\author{Natsuki H.~\textsc{Hayatsu}\altaffilmark{14,15}}
\altaffiltext{14}{Department of Physics, Graduate School of Science, The University of Tokyo, 7-3-1 Hongo, Bunkyo, Tokyo 113-0033, Japan}
\altaffiltext{15}{European Southern Observatory, Karl-Schwarzschild-Str.~2, D-85748 Garching, Germany}

\author{Rob J.~\textsc{Ivison}\altaffilmark{15,6}}

\author{Tadayuki \textsc{Kodama}\altaffilmark{16}}
\altaffiltext{16}{Astronomical Institute, Tohoku University, 6-3 Aramaki, Aoba, Sendai, Miyagi 980-8578, Japan}

\author{Haruka \textsc{Kusakabe}\altaffilmark{17,18}}
\altaffiltext{17}{Department of Astronomy, Graduate School of Science, The University of Tokyo, 7-3-1 Hongo, Bunkyo, Tokyo 113-0033, Japan}
\altaffiltext{18}{Observatoire de Gen\'eve, Universit\'e de Gen\'eve, 51 chemin de P\'egase, 1290 Versoix, Switzerland}

\author{Tohru \textsc{Nagao}\altaffilmark{19}}
\altaffiltext{19}{Research Center for Space and Cosmic Evolution, Ehime University, 2-5 Bunkyo-cho, Matsuyama, Ehime 790-8577, Japan}

\author{Masami \textsc{Ouchi}\altaffilmark{3,10,20}}
\altaffiltext{20}{Kavli Institute for the Physics and Mathematics of the Universe (Kavli IPMU), WPI, The University of Tokyo, Kashiwa, Chiba 277-8583, Japan}

\author{Wiphu \textsc{Rujopakarn}\altaffilmark{10, 21, 22}}
\altaffiltext{21}{Department of Physics, Faculty of Science, Chulalongkorn University, 254 Phayathai Road, Pathumwan, Bangkok 10330, Thailand}
\altaffiltext{22}{National Astronomical Research Institute of Thailand (Public Organization), Don Kaeo, Mae Rim, Chiang Mai 50180, Thailand}

\author{Ken-ichi \textsc{Tadaki}\altaffilmark{3}}

\author{Yoichi \textsc{Tamura}\altaffilmark{23}}
\altaffiltext{23}{Division of Particle and Astrophysical Science, Nagoya University, Furocho, Chikusa, Nagoya 464-8602, Japan}

\author{Yoshihiro \textsc{Ueda}\altaffilmark{24}}
\altaffiltext{24}{Department of Astronomy, Kyoto University, Kyoto 606-8502, Japan}

\author{Hideki \textsc{Umehata}\altaffilmark{25,1}}
\altaffiltext{25}{RIKEN Cluster for Pioneering Research, 2-1 Hirosawa, Wako-shi, Saitama 351-0198, Japan}

\author{Wei-Hao \textsc{Wang}\altaffilmark{26,27}}
\altaffiltext{26}{Academia Sinica Institute of Astronomy and Astrophysics (ASIAA), No.~1, Sec.~4, Roosevelt Rd., Taipei 10617, Taiwan}
\altaffiltext{27}{Canada-France-Hawaii Telescope (CFHT), 65-1238 Mamalahoa Hwy., Kamuela, HI 96743, USA}



\maketitle

\begin{abstract}
We make use of the ASAGAO, deep 1.2 mm continuum observations of a 26 arcmin$^2$ region in the GOODS-South field obtained with ALMA, to probe dust-enshrouded star formation in  $K$-band selected (i.e., stellar mass selected) galaxies, which are drawn from the ZFOURGE catalog. Based on the ASAGAO combined map, which was created by combining ASAGAO and ALMA archival data in the GOODS-South field, we find that 24 ZFOURGE sources have 1.2 mm counterparts with a signal-to-noise ratio $>$ {4.5} (1$\sigma\simeq$ 30--70 $\mu$Jy beam$^{-1}$ at 1.2 mm). 
Their median redshift is estimated to be $z_\mathrm{median}=$ {2.38 $\pm$ 0.14}.
They generally follow the tight relationship of the stellar mass versus star formation rate (i.e., the main sequence of star-forming galaxies). ALMA-detected ZFOURGE sources exhibit systematically larger infrared (IR) excess (IRX $\equiv L_\mathrm{IR}/L_\mathrm{UV}$) compared to ZFOURGE galaxies without ALMA detections even though they have similar redshifts, stellar masses, and star formation rates. This implies the consensus stellar-mass versus IRX relation, which is known to be tight among rest-frame-UV-selected galaxies, can not fully predict the ALMA detectability of stellar-mass-selected galaxies. We find that ALMA-detected ZFOURGE sources are the main contributors to the cosmic IR star formation rate density at $z$ = 2--3. 
\end{abstract}

\section{Introduction} \label{sec:intro}

Recent studies have revealed the evolution of the cosmic star formation rate density (SFRD) as a function of redshift based on various wavelengths (e.g., \cite{madau2014, bouwens2015, bouwens2016}, and references therein). The roles of dust-obscured star-formation in star-forming galaxies at redshift $z\simeq$ 1--3 and beyond are one of the central issues, because the majority of star-forming galaxies at $z\simeq$ 1--3, where the cosmic star formation activity peaks, are dominated by dust-enshrouded star-formation. 


At (sub-)millimeter wavelengths, several studies have found bright sub-millimeter galaxies (SMGs) whose observed flux densities are larger than a few mJy at (sub-)millimeter wavelengths (i.e., $\sim$ 850 $\mu$m--1 mm) in blank-field bolometer surveys (e.g., \cite{smail1997, hughes1998, barger1998, blain2002, greve2004, weiss2009, scott2010, hatsukade2011, casey2013, umehata2014}, and references therein). The fact that (sub-)millimeter flux densities are almost constant at $z>1$ for galaxies with a given infrared (IR) luminosity (i.e., the negative $k$-correction -- e.g., \cite{blain1996}) makes it efficient to study dust-obscured star-formation activity at high redshift and the extreme star-formation rates (SFRs) of SMGs [a few 100—-1000 $M_\odot$ yr$^{-1}$, modulo expectations for and observations of the stellar initial mass function (IMF) in starburst environments -- \cite{papadopoulos2011, zhang2018}] make them non-negligible contributors to the cosmic SFRD (e.g., \cite{hughes1998, casey2013, wardlow2011, swinbank2014}).

Deep (sub-)millimeter-wave surveys, using 
the James Clerk Maxwell Telescope/ Submillimeter Common-Use Bolometer Array 2
(SCUBA2; \cite{holland2013}), AzTEC \citep{wilson2008} on Atacama Submillimeter Telescope Experiment (ASTE; \cite{ezawa2004, ezawa2008}), LABOCA \citep{siringo2009} on Atacama Pathfinder EXperiment (APEX; \cite{gusten2006})\textcolor{blue}{,} 
\textit{Herschel}/Spectral and Photometric Imaging Receiver (SPIRE; \cite{griffin2010}) and so on, play essential roles in revealing the contributions of dust-obscured star formation activities (e.g., \cite{elbaz2011, burgarella2013}), but their limited angular resolution does not allow us to measure far-IR fluxes of individual sources if we go down to luminous IR galaxy (LIRG) class sources [i.e., IR luminosity $(L_\mathrm{IR})\sim$ 10$^{11}$ $L_\odot$]. Indeed, the contribution of these ``classical'' SMGs ($L_\mathrm{IR}\sim$ 10$^{12}$--10$^{13}$ $L_\odot$) to the integrated extragalactic background light is not so large ($\sim$ 20--40\% at 850 $\mu$m and $\sim$ 10--20\% at 1.1 mm; e.g., \cite{eales1999, coppin2006, weiss2009, hatsukade2011, scott2012}). This means that the bulk of dust-obscured star formation activities in the universe remained unresolved due to the confusion limit of single-dish telescopes.

Even with single-dish telescopes, we can access the fainter (sub-)millimeter population (i.e., observed flux densities $S_\mathrm{obs} \lesssim$ 1 mJy) using gravitational magnification by lensing clusters or stacking analysis (e.g., \cite{knudsen2008, geach2013, coppin2015}). However, in lensed object surveys, the effective sensitivity comes at the cost of a reduced survey volume\footnote{\citet{knudsen2008} suggest that the effective (source-plane) area within sufficient magnification to detect fainter (sub-)millimeter populations is only $\sim$ 0.1 arcmin$^2$ for a typical rich cluster.}, which increases the cosmic variance uncertainty (e.g., \cite{robertson2014}). The stacking technique is a useful way to obtain the average properties of less-luminous populations, but individual source properties have remained unexplored. Therefore, more sensitive observations with higher angular resolution are needed.

The advent of the Atacama Large Millimeter/sub-millimeter Array (ALMA), which offers high sensitivity and angular resolution capabilities, has allowed the fainter (sub-)millimeter population to be revealed below the confusion limit of single-dish telescopes. 
For instance, 
the ALMA follow-up observation of the LABOCA Extended Chandra Deep Field South surveys (ALESS; e.g., \cite{hodge2013, swinbank2014, dacunha2015}) have yielded detections of faint submillimeter sources. Archival ALMA data has also been exploited to find many faint (sub-)millimeter sources 
(e.g., \cite{hatsukade2013, fujimoto2016, oteo2016}). 
%
ALMA has also been used to obtain ``confusion-free'', deep contiguous maps 
in SXDF-UDS-CANDELS ($\sim$2 arcmin$^2$,
\cite{tadaki2015,kohno2016,hatsukade2016,wang2016}) 
and (proto-)cluster fields including Hubble Frontier Fields ($\sim$4 arcmin$^2$ per cluster, e.g., \cite{gonzalez2017, munoz2018}) and SSA22 ($\sim$6 to 20 arcmin$^2$, \cite{umehata2017, umehata2018}). 
Tiered ALMA deep surveys with a ``wedding-cake'' approach have been conducted 
in \textit{Hubble} Ultra-Deep Field (HUDF, $\sim$1-4 arcmin$^2$, \cite{aravena2016,walter2016,rujopakarn2016,dunlop2017,gonzalez2020}) and GOODS-S ($\sim$26 arcmin$^2$,  \cite{ueda2018,hatsukade2018}, and $\sim$69 arcmin$^2$, \cite{franco2018,franco2020}). 


Faint (sub-)millimeter sources uncovered by these ALMA observations 
tend to preferentially have large stellar masses ($\gtrsim$ 10$^{10}$ $M_\odot$, \cite{tadaki2015, aravena2016, bouwens2016, dunlop2017}). In fact, a tight correlation between the stellar masses and the infrared excesses or IRXs, defined as a ratio of IR luminosity to UV luminosity ($L_\mathrm{IR}/L_\mathrm{UV}$), has been proposed (e.g., \cite{bouwens2016, fudamoto2017, koprowski2018}), mainly based on the ALMA fluxes of rest-frame-UV-selected galaxies such as Lyman break galaxies (LBGs). 
However, it is not entirely clear if the stellar mass is the unique parameter to predict IRXs in galaxies, and whether such a trend can be applicable to other types of galaxies such as rest-frame-optical-selected galaxies. It is also intriguing to see if there are low-mass galaxies with an elevated IRX or high-mass galaxies with a low IRX. Currently, the number of galaxies with both stellar-mass and IRX measurements using ALMA is still insufficient to address these questions.

\medskip


Here, we present millimeter-wave properties of $K$-band selected galaxies 
in the \texttt{FourStar} galaxy evolution survey (ZFOURGE)\footnote{http://zfourge.tamu.edu/} catalog \citep{straatman2016} by exploiting the ALMA twenty-Six Arcmin$^2$ survey of GOODS-S At One-millimeter (ASAGAO; Project ID: 2015.1.00098.S, PI: K.~Kohno)\footnote{https://sites.google.com/view/asagao26/}, 
one of the tiered ALMA deep surveys in HUDF/GOODS-S, 
to constrain dust-enshrouded star-forming properties of mass-selected galaxies and 
assess their contribution to the cosmic SFRD. 
The ZFOURGE catalog contains 30,911 $K$-band selected galaxies over 128 arcmin$^2$ in the {\it Chandra} Deep Field South, which fully includes the ASAGAO field, with 5$\sigma$ limiting AB magnitude of $K_s$ = 26.0 to 26.3 at the 80\% and 50\% completeness levels (with masking), respectively. There are $\simeq$ 3,283 ZFOURGE sources within the ASAGAO field. Thanks to the high resolution of the ALMA mosaic image ($\simeq$ 0''.5, see Section \ref{sec:data} for details), we can select ALMA-detected $K$-band sources reliably to constrain their dusty star-formation properties. 

This paper is structured as follows. Section \ref{sec:data} presents our ALMA observations and the source identifications. Then, we describe our strategy to obtain Spectral Energy Distribution (SED) fits in Section \ref{sec:SEDfit}, and we discuss their derived physical properties in Section \ref{sec:multi-w prop}. In Section \ref{sec:SFRD}, we explain the contribution of $K$-band-detected ASAGAO sources to the cosmic SFRD. Section \ref{sec:summary} presents our conclusions. Throughout this paper, we assume a $\Lambda$ cold dark matter cosmology with $\Omega_\mathrm{M}$ = 0.3, $\Omega_\mathrm{\Lambda}$ = 0.7, and $H_0$ = 70 km s$^{-1}$ Mpc$^{-1}$. All magnitudes are given according to the AB system. We adopt the Chabrier IMF \citep{chabrier2003} in this paper.

\section{ZFOURGE sources with ALMA counterparts} \label{sec:data}

\subsection{ALMA Band-6 data}

\begin{figure*}[t!]
\begin{center}
\includegraphics[width=120mm,angle=270]{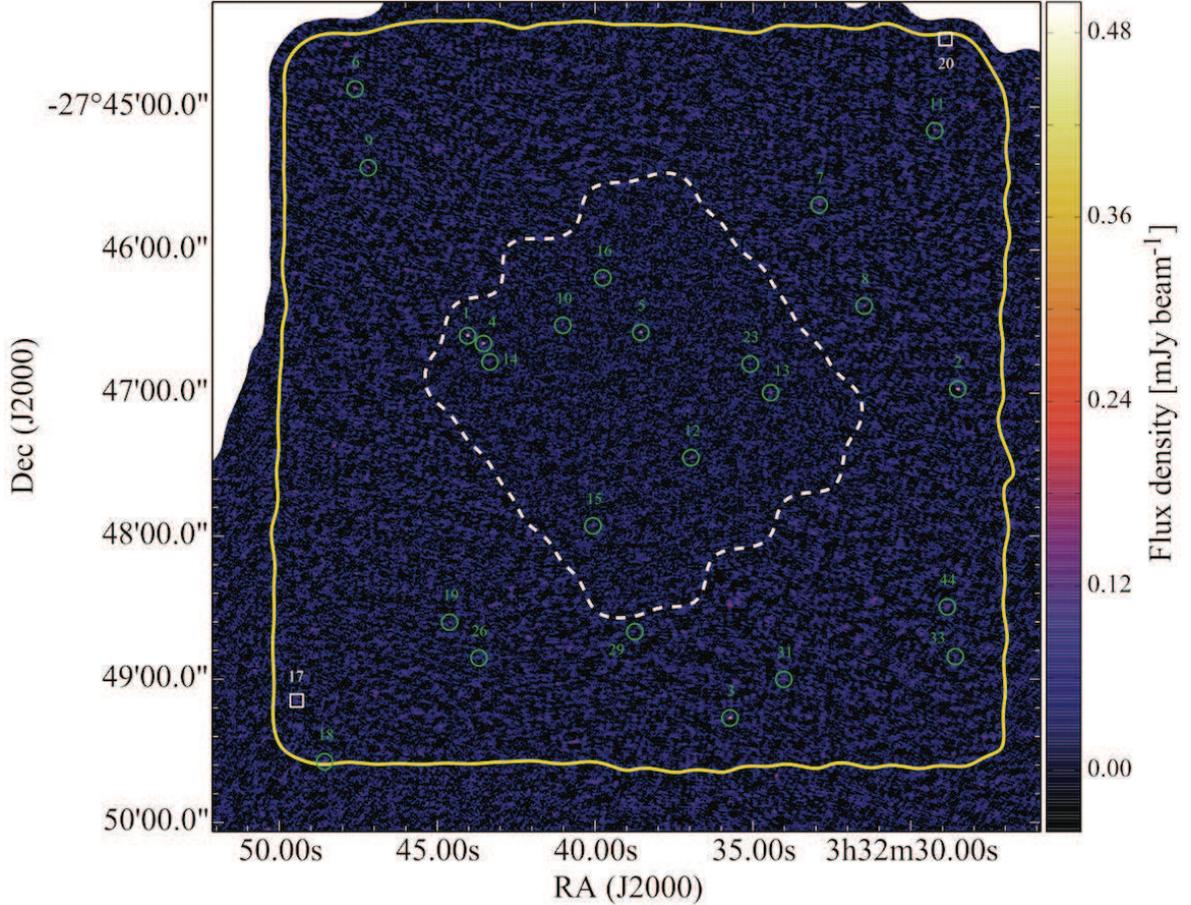}
\end{center}
\caption{ASAGAO 1.2-mm continuum map of GOODS-S. ASAGAO original data, HUDF data \citep{dunlop2017}, and a part of GOODS-ALMA data \citep{franco2018} are combined. In this paper, we only consider the ASAGAO field indicated by the yellow solid line ($\sim$ 5$^{\prime}$ $\times$ 5$^{\prime}$). The white dashed line indicates the area covered by \citet{dunlop2017}. The green symbols indicate 
{24} ASAGAO continuum sources with $K$-band counterparts (see Section \ref{subsec:ID}). 
Two white squares show the positions of secure (S/N $>$ 5.5) ASAGAO sources without ZFOURGE counterparts, which have been reported in a separate paper \citep{yamaguchi2019}.
\label{fig:area}}
\end{figure*}

In this paper, we use the ALMA data obtained by ASAGAO. As presented in \citet{hatsukade2018}, the 26 arcmin$^2$ map of the ASAGAO field was obtained at 1.14 mm and 1.18 mm (two tunings) to cover a wider frequency range, whose central wavelength was 1.16 mm. 
In addition to the original ASAGAO data, we also included ALMA archival data of the same field (Project ID: 2015.1.00543.S, PI: D.~Elbaz and Project ID: 2012.1.00173.S, PI: J.~S.~Dunlop) to improve the sensitivity. The data were imaged with the Common Astronomy Software Applications package (\texttt{CASA}; \cite{mcmullin2007}) version 5.1.1, but calibration was done with the version 4.7.2. The maps were processed with the \texttt{CLEAN} algorithm \citep{hogdom1974} with the task \texttt{tclean}. Details of the data analysis are given in \citet{hatsukade2018}. The combined map reached typical rms noise of 30--70 $\mu$Jy beam$^{-1}$ with a synthesized beam of 0$^{\prime\prime}$.59 $\times$ 0$^{\prime\prime}$.53 (PA = $-$83$^\circ$). Note that the typical sensitivity is calculated within the area covered by ASAGAO (i.e., the region enclosed by the yellow solid line shown in Figure \ref{fig:area}).

\subsection{ALMA counterparts identification} \label{subsec:ID}

Since it has been reported that astrometric corrections are necessary for sources catalogued using HST and ZFOURGE images in GOODS-S (e.g., \cite{rujopakarn2016,dunlop2017,franco2018}), the ZFOURGE source coordinates were 
corrected by $-$0$^{\prime\prime}$.086 in right ascension and $+$0$^{\prime\prime}$.282 in declination, which is calibrated by the positions of stars in the Gaia Data Release 1 catalog \citep{gaia2016} within the ASAGAO field. 
We then measure ALMA flux densities of ZFOURGE sources.
Although \citet{bouwens2016} consider a S/N threshold of 2.0 to search for ALMA counterparts of LBGs, we adopt a more conservative threshold of S/N $=$ 4.5. 
We extracted 45 positive sources and 9 negative sources (i.e., false detections) with S/N $>$ 4.5. Therefore, the ratio between the number of negative sources and positive sources is 0.2.

For point-like ZFOURGE sources, we allow the positional offsets between ZFOURGE and ALMA positions of less than $0^{\prime\prime}.5$, which is comparable with the synthesized beam of the combined ALMA map. Considering the number of ZFOURGE sources within the ASAGAO field ($\sim$ 3,000), the likelihood of random coincidence is estimated to be 0.03 (this likelihood is often called the $p$-value; \cite{downes1986}). In the case that a counterpart is largely extended, we allow a larger positional offset, up to half-light radius of $K_s$-band emission. We exclude ZFOURGE sources with ``\texttt{use flag} $=$ 0'' (e.g., sources with low S/N at $K$-band or catastrophic SED fits; see \cite{straatman2016}, for details) in order to prevent mismatching. {When we apply the same procedure to the negative values of the ALMA map, we find that {no negative sources with an S/N $\lesssim$ $-$4.5 show chance coincidence.} This coincidence rate is comparable with the estimated value by \citet{casey2018}}.

\begin{figure*}
\begin{center}
\includegraphics[width = 160mm]{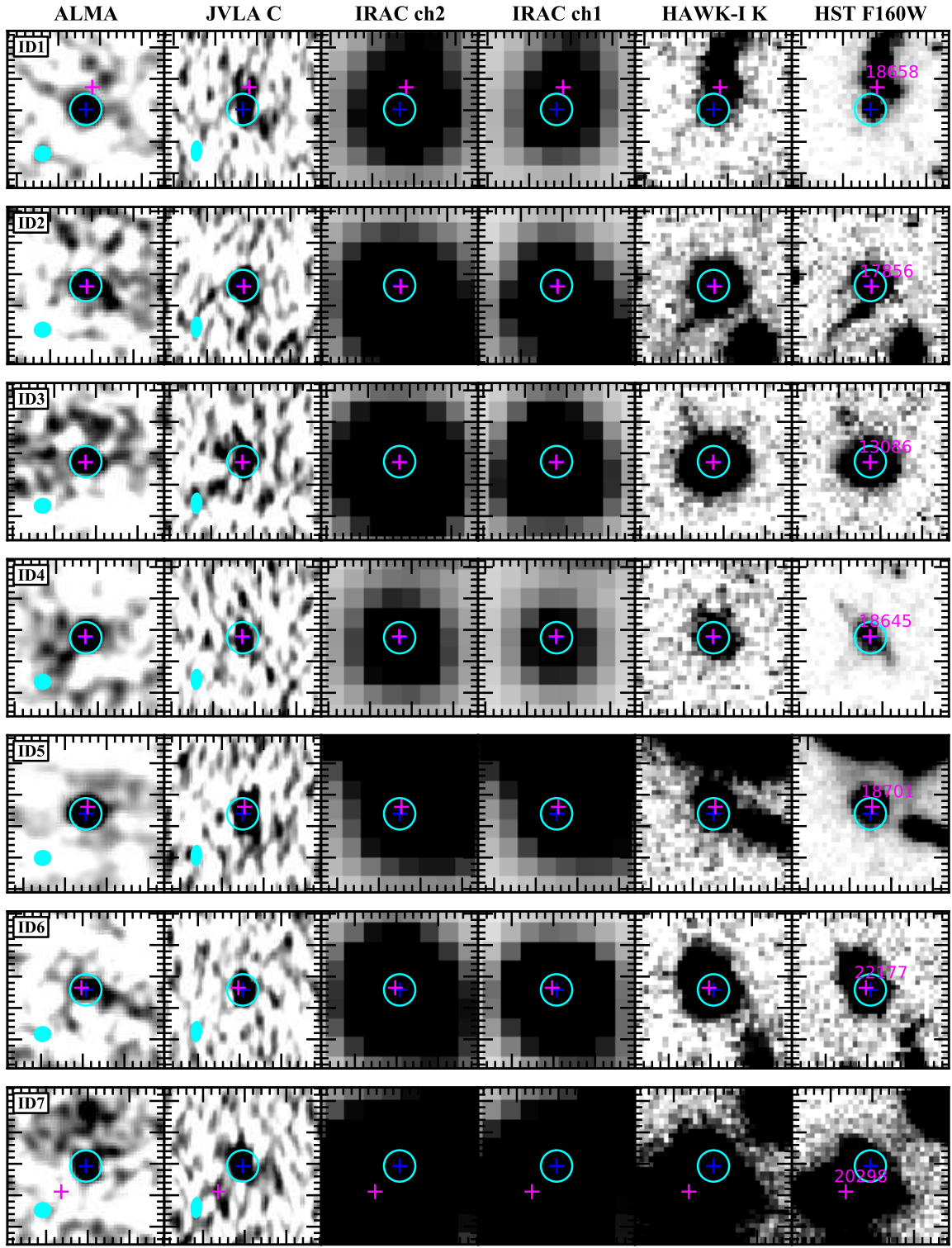}
\end{center}
\caption{Multi-wavelength images of ASAGAO sources with $K$-band counterparts. From left to right: ALMA 1.2 mm, JVLA 6 GHz (C-band), \textit{Spitzer} IRAC/4.5 $\mu$m, IRAC/3.6 $\mu$m, VLT HAWK-I/$K_s$, and HST WFC3/$F160W$ images. The field of view is 5$^{\prime\prime}$ $\times$ 5$^{\prime\prime}$. Blue and magenta crosses mark the ALMA positions and ZFOURGE positions, respectively. Cyan circles are 1$^{\prime\prime}$ apertures. The synthesized beams of ALMA and JVLA are expressed as cyan ellipses. 
ZFOURGE source IDs are shown in the HST/$F160W$ images (in magenta). 
\label{fig:postagestamp}}
\end{figure*}
\begin{figure*}
\begin{center}
\includegraphics[width = 160mm]{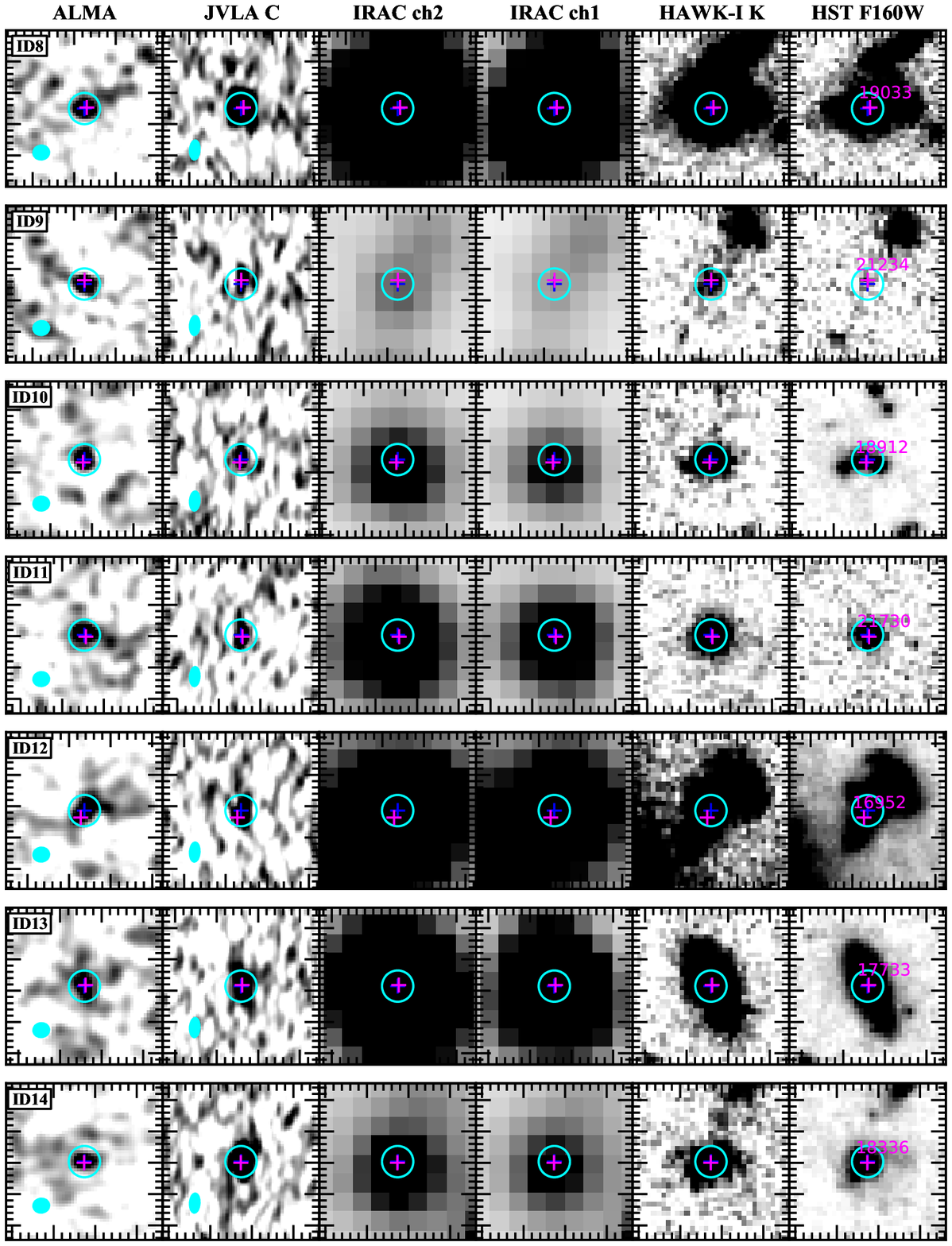}
\\Figure \ref{fig:postagestamp} (Continued.)
\end{center}
\end{figure*}
\begin{figure*}
\begin{center}
\includegraphics[width = 160mm]{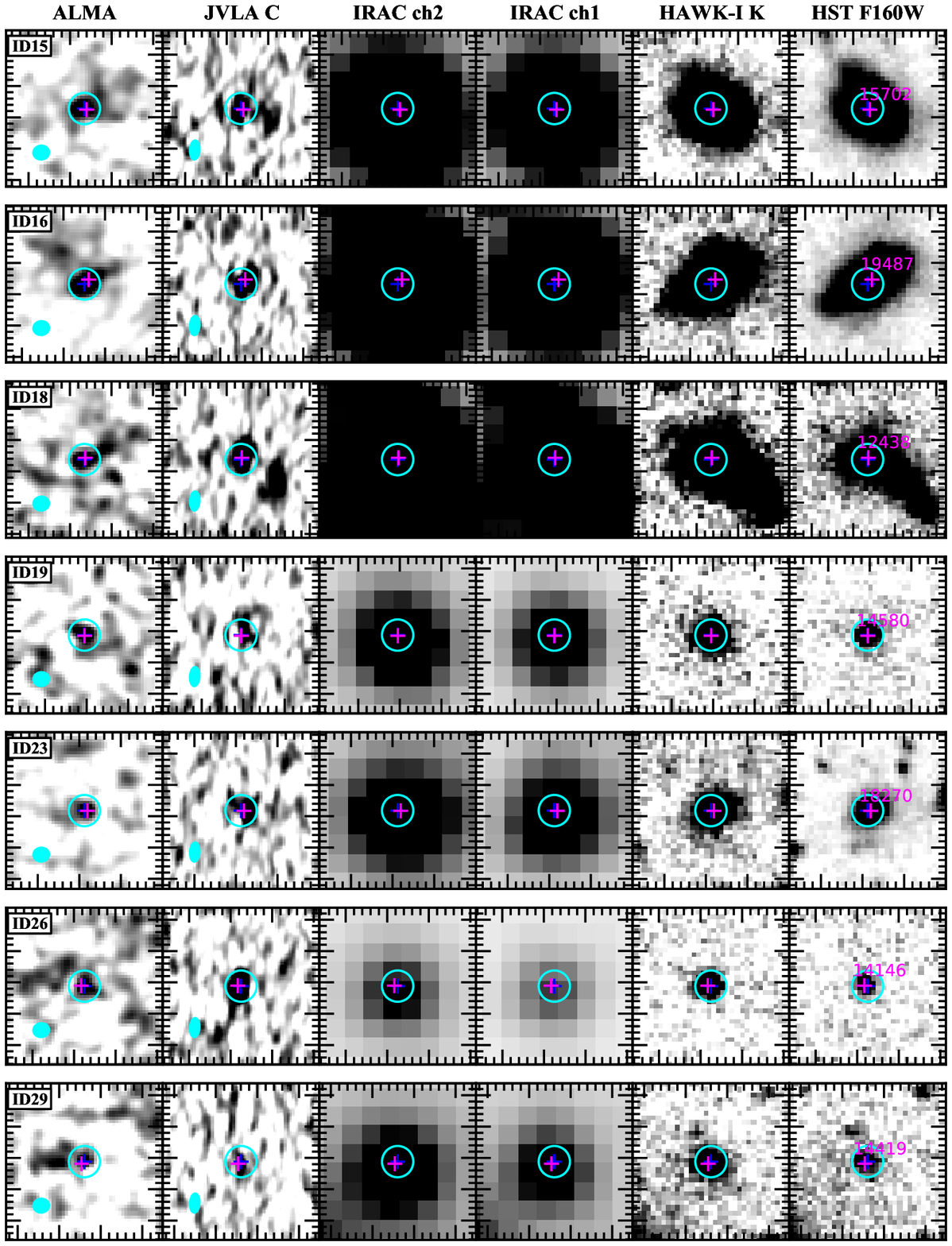}
\\Figure \ref{fig:postagestamp} (Continued.)
\end{center}
\end{figure*}
\begin{figure*}
\begin{center}
\includegraphics[width = 160mm]{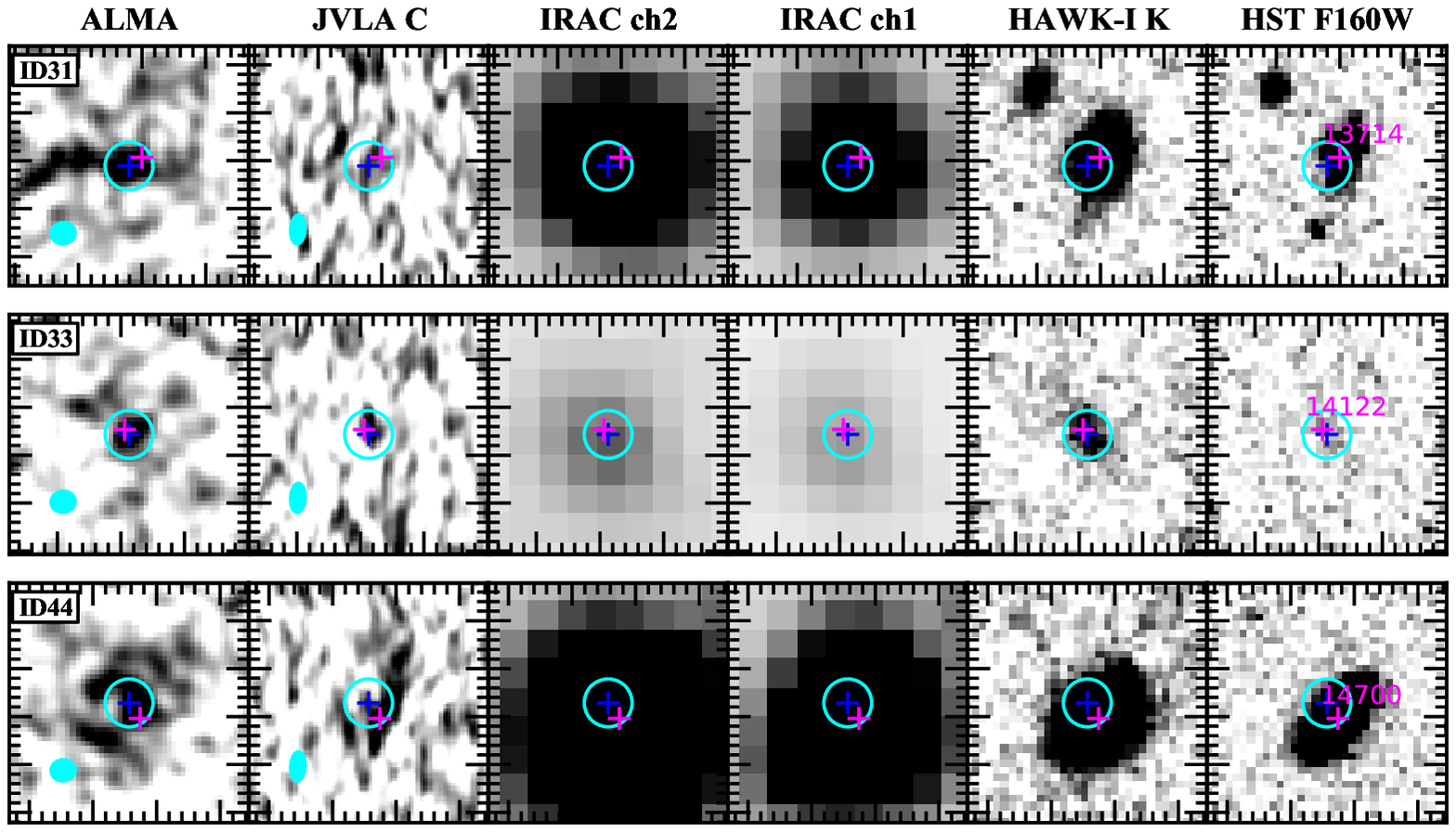}
\\Figure \ref{fig:postagestamp} (Continued.)
\end{center}
\end{figure*}

Flux measurements in the ALMA map were performed at the position of ZFOURGE sources considering positional offset as explained above. We consider the flux-boosting effect by calculating the ratio between input and output integrated flux densities of inserted 30,000 artificial sources into the signal map (see \cite{hatsukade2018}, for details).  The effect of flux boosting for the sources with S/N $>$ 4.5 is $\lesssim$ 15\% \citep{hatsukade2018}, which is comparable with previous studies.

Finally, we identify {24} ZFOURGE sources that have ALMA counterparts (hereafter, we define them as ASAGAO sources). 
Note that two ALMA sources without ZFOURGE source associations, or ``NIR-dark ALMA sources'', have been reported in a separate paper (\cite{yamaguchi2019}). 
In Table \ref{tab:counterpart}, we summarize ALMA fluxes of ZFOURGE sources 
in order of ALMA peak S/N. As shown in Table \ref{tab:counterpart}, some ASAGAO sources show larger $p$-value than the traditional threshold of $p<0.05$ (e.g., \cite{biggs2011,casey2013}). {We remove these ASAGAO sources with $p>0.05$ (i.e., ID1 and ID7) from our conclusions presented in Section \ref{sec:multi-w prop} and Section \ref{sec:SFRD} to prevent for miss identifications.} We show the positions of ASAGAO sources and their multi-wavelength postage stamps in Figure \ref{fig:area} and Figure~\ref{fig:postagestamp}, respectively. 

\citet{ueda2018} and \citet{fujimoto2018} also use ASAGAO data. In the tables of Appendix \ref{sec:correspond}, we present the correspondence of their ID to ASAGAO ID, which is presented in this paper.
%
%
We also cross-matched the ASAGAO sources with 1.3 mm sources of HUDF \citep{dunlop2017}, 1.1 mm sources of GOODS-ALMA \citep{franco2018}, 1.2 mm sources of ASPECS \citep{aravena2016}, and 870 $\mu$m sources obtained by \citet{cowie2018}. The results of cross-matching are presented in Table \ref{tab:ID_goods-s} in Appendix \ref{sec:correspond}. 

\begin{figure}[t!]
\begin{center}
\includegraphics[width=80mm]{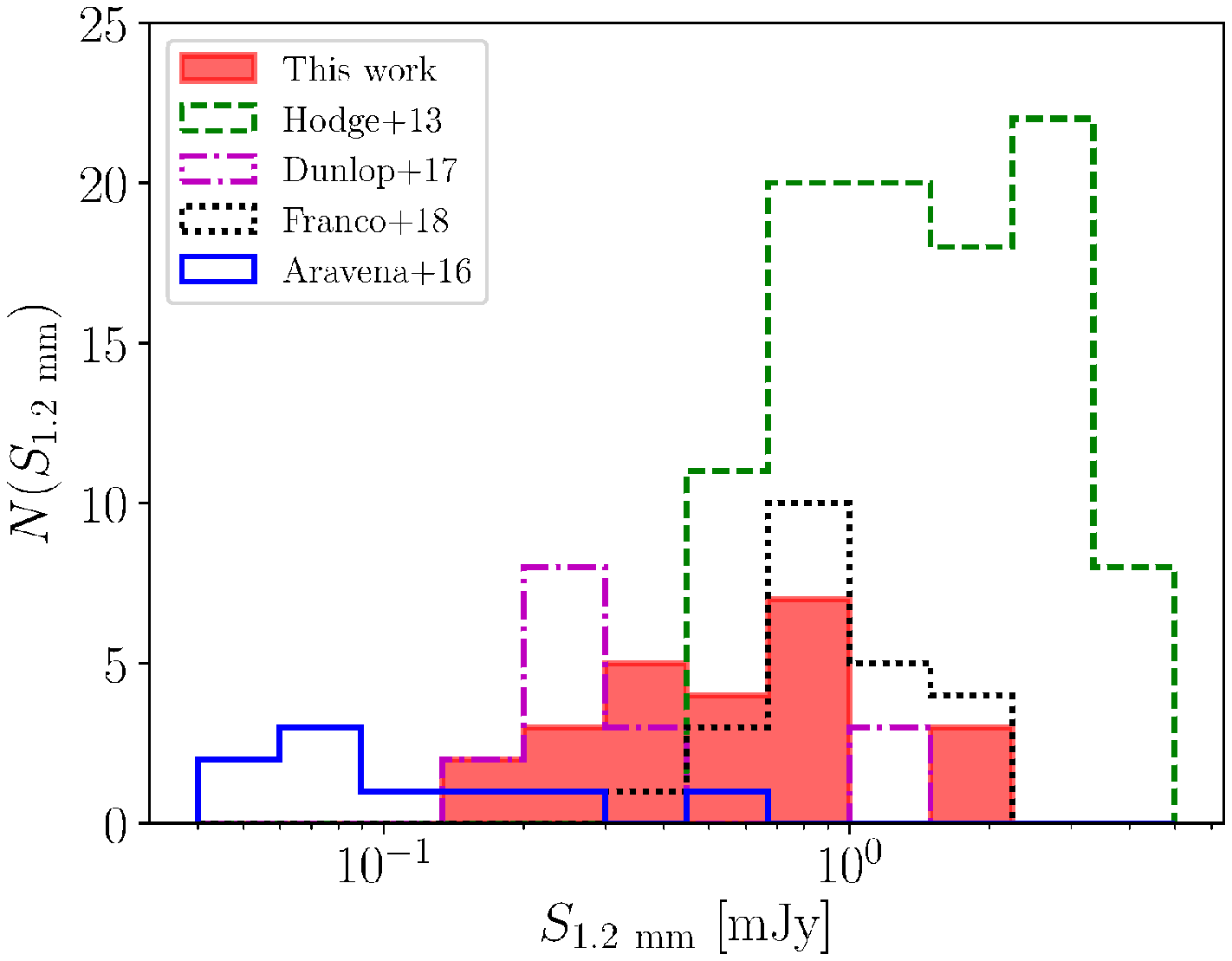}
\end{center}
\caption{Flux-density distribution of ASAGAO sources (red-shaded region). The flux-density distributions of other ALMA continuum source with optical/near-IR counterparts in ALESS \citep{hodge2013}, HUDF \citep{aravena2016, dunlop2017}, and GOODS-S \citep{franco2018} are also shown. 
\label{fig:flux_hist}}
\end{figure}

\subsection{Observed flux densities at 1.2 mm}
In Figure \ref{fig:flux_hist}, we plot the histogram of observed flux densities of ASAGAO sources at 1.2 mm. As a comparison, we also show the histograms of observed flux densities obtained by ALESS \citep{hodge2013,dacunha2015}, HUDF \citep{dunlop2017}, GOODS-ALMA \citep{franco2018}, and ASPECS \citep{aravena2016}. Note that ALESS sources, HUDF sources, and GOODS-ALMA sources were observed at 870 $\mu$m, 1.3 mm, and 1.1 mm, respectively. Therefore, we converted these flux densities to 1.2-mm flux densities with the assumption of a modified blackbody with a dust emissivity index of 1.5 and dust temperature of 35 K\footnote{for example, $S_\mathrm{1.2\ mm}/S_\mathrm{870\ \mu m} =0.44,\  S_\mathrm{1.2\ mm}/S_\mathrm{1.3\ mm} =1.26$, and $S_\mathrm{1.2\ mm}/S_\mathrm{1.1\ mm} = 0.79$ at $z$ = 2.83, 2.04, and 2.54 (median redshifts of ALESS, HUDF, and GOODS-ALMA sources)}.

{Figure \ref{fig:flux_hist} shows that ASAGAO sources tend to have fainter flux densities ($S_\mathrm{1.2\ mm}\lesssim$ 1 mJy) than most of the ALESS sources ($S_\mathrm{1.2\ mm}\gtrsim$ 1 mJy). Although recent ALMA contiguous surveys focusing on stellar mass selected sources (e.g., \cite{aravena2016, dunlop2017, franco2018}) also suggest that their samples tend to have the flux densities of $S_\mathrm{1.2\ mm}\lesssim$ 1 mJy, we provide the largest number of  stellar mass selected sources with 1.2 mm flux densities.
}

\begin{sidewaystable*}[]
\footnotesize{
\caption{ZFOURGE sources with ALMA counterparts \label{tab:counterpart}}
\begin{center}
\begin{tabular}{ccccccccccccc}
\hline\hline
{ID} & {RA$_\mathrm{ZFOURGE}$\footnotemark[a]} & {Dec.$_\mathrm{ZFOURGE}$\footnotemark[a]}
& {ID} & {$S_\mathrm{ALMA}$} & {S/N$_\mathrm{peak}$} & 
{RA$_\mathrm{ALMA}$} & {Dec.$_\mathrm{ALMA}$} 
&  {{$|\Delta_\mathrm{offset}|$}} & {{$p$-value}}
& {$z_\mathrm{photo}$} & {$z_\mathrm{spec}$} 
& {{\it Chandra}}\\
{(ZFOURGE)} & {(deg.)} & {(deg.)} & {(ASAGAO)} &
{(mJy)} & {} & {(deg.)} & {(deg.)} & {(arcsec)}  & {} & {}  
& {} & {counterpart?}\\
(1) & (2) & (3) & (4) & (5) & (6) & (7) & (8) & (9) & (10) & (11) & (12) & (13)\\
\hline
18658 & 53.18341 & $-$27.77646 & 1 & 0.985$\pm$0.036 & 25.995 & 53.18348 & $-$27.77666 & 0.735 & 0.0578 & 2.83$^{+0.07}_{-0.08}$ & -- & Y\\
17856 & 53.11880 & $-$27.78289 & 2 & 1.973$\pm$0.075 & 25.625 & 53.11881 & $-$27.78288 & 0.048 & 0.0003 & 2.38$^{+0.17}_{-0.10}$ & -- & Y\\
13086 & 53.14885 & $-$27.82119 & 3 & 1.748$\pm$0.070 & 24.008 & 53.14885 & $-$27.82119 & 0.021 & 0.0 & 2.58$^{+0.04}_{-0.04}$ & 2.582 & Y\\
18645 & 53.18137 & $-$27.77756 & 4 & 0.906$\pm$0.041 & 21.045 & 53.18137 & $-$27.77757 & 0.044 & 0.0002 & 2.92$^{+0.06}_{-0.06}$ & -- &  \\
18701 & 53.16061 & $-$27.77622 & 5 & 0.735$\pm$0.039 & 18.101 & 53.16063 & $-$27.77628 & 0.228 & 0.0057 & 2.61$^{+0.07}_{-0.05}$ & 2.543\footnotemark[b] & Y\\
22177 & 53.19835 & $-$27.74788 & 6 & 0.922$\pm$0.074 & 12.421 & 53.19830 & $-$27.74790 & 0.153 & 0.0026 & 1.93$^{+0.04}_{-0.03}$ & -- & Y\\
20298 & 53.13735 & $-$27.76163 & 7 & 0.778$\pm$0.086 & 8.785 & 53.13710 & $-$27.76141 & 1.124 & 0.13 & 0.52$^{+0.02}_{-0.01}$ & 0.523 &  \\
19033 & 53.13112 & $-$27.77319 & 8 & 0.610$\pm$0.072 & 8.654 & 53.13115 & $-$27.77320 & 0.084 & 0.0008 & 2.22$^{+0.03}_{-0.03}$ & 2.225 & Y\\
21234 & 53.19656 & $-$27.75704 & 9 & 0.457$\pm$0.055 & 8.575 & 53.19656 & $-$27.75708 & 0.123 & 0.0017 & 2.46$^{+0.05}_{-0.05}$ & -- &  \\
18912 & 53.17092 & $-$27.77547 & 10 & 0.261$\pm$0.031 & 8.550 & 53.17091 & $-$27.77544 & 0.099 & 0.0011 & 2.36$^{+0.10}_{-0.11}$ & -- &  \\
21730 & 53.12185 & $-$27.75278 & 11 & 0.635$\pm$0.078 & 8.506 & 53.12186 & $-$27.75277 & 0.071 & 0.0006 & 2.01$^{+0.06}_{-0.04}$ & -- & Y\\
16952 & 53.15405 & $-$27.79093 & 12 & 0.376$\pm$0.049 & 7.378 & 53.15401 & $-$27.79087 & 0.251 & 0.0069 & 1.88$^{+0.04}_{-0.03}$ & 1.317\footnotemark[b] &  \\
17733 & 53.14349 & $-$27.78328 & 13 & 0.400$\pm$0.053 & 7.227 & 53.14351 & $-$27.78329 & 0.05 & 0.0003 & 1.62$^{+0.04}_{-0.05}$ & 1.415\footnotemark[b] &  \\
18336 & 53.18053 & $-$27.77972 & 14 & 0.238$\pm$0.035 & 7.178 & 53.18053 & $-$27.77971 & 0.038 & 0.0002 & 2.67$^{+0.11}_{-0.15}$ & -- & Y\\
15702 & 53.16692 & $-$27.79882 & 15 & 0.416$\pm$0.064 & 6.637 & 53.16694 & $-$27.79881 & 0.082 & 0.0007 & 1.93$^{+0.03}_{-0.03}$ & 1.998 & Y\\
19487 & 53.16558 & $-$27.76987 & 16 & 0.488$\pm$0.065 & 6.491 & 53.16562 & $-$27.76991 & 0.194 & 0.0041 & 1.61$^{+0.08}_{-0.06}$ & 1.551\footnotemark[b] & Y\\
12438 & 53.20235 & $-$27.82627 & 18 & 0.975$\pm$0.172 & 5.803 & 53.20236 & $-$27.82629 & 0.063 & 0.0004 & 1.07$^{+0.03}_{-0.03}$ & -- & Y\\
14580 & 53.18585 & $-$27.81004 & 19 & 0.387$\pm$0.073 & 5.659 & 53.18585 & $-$27.81004 & 0.024 & 0.0001 & 2.81$^{+0.10}_{-0.10}$ & 2.593 & Y\\
18270 & 53.14617 & $-$27.77995 & 23 & 0.182$\pm$0.037 & 5.360 & 53.14620 & $-$27.77995 & 0.096 & 0.001 & 2.61$^{+0.05}_{-0.06}$ & -- & Y\\
14146 & 53.18201 & $-$27.81420 & 26 & 0.222$\pm$0.052 & 4.923 & 53.18198 & $-$27.81420 & 0.107 & 0.0013 & 2.41$^{+0.18}_{-0.14}$\footnotemark[c] & -- & Y\\
14419 & 53.16144 & $-$27.81116 & 29 & 0.197$\pm$0.046 & 4.835 & 53.16141 & $-$27.81114 & 0.115 & 0.0015 & 2.77$^{+0.11}_{-0.10}$ & -- & Y\\
13714 & 53.14167 & $-$27.81665 & 31 & 0.733$\pm$0.158 & 4.714 & 53.14175 & $-$27.81670 & 0.328 & 0.0118 & 2.53$^{+0.09}_{-0.10}$ & -- & Y\\
14122 & 53.11914 & $-$27.81402 & 33 & 0.318$\pm$0.079 & 4.701 & 53.11911 & $-$27.81405 & 0.136 & 0.002 & 3.32$^{+0.44}_{-0.45}$ & -- &  \\
14700 & 53.12011 & $-$27.80834 & 44 & 1.768$\pm$0.447 & 4.546 & 53.12018 & $-$27.80825 & 0.401 & 0.0175 & 1.83$^{+0.05}_{-0.05}$ & -- &  \\
\hline
\end{tabular}
\end{center}
\begin{tabnote}
{\bf Notes.} {ZFOURGE sources with ALMA counterpart in order of ALMA S/N. (1) ZFOURGE ID. (2) and (3) ZFOURGE position. (4) ASAGAO ID. (5) Spatially integrated ALMA flux density (de-boosted). (6) ALMA peak S/N. (7) and (8) ASAGAO position. {(9) The positional offset between ALMA and ZFOURGE. (10) The $p$-Values for each ASAGAO source.} (11) The photometric redshift. (12) The spectroscopic redshift. {(13) Based on cross-matching with the Chandra catalog \citep{luo2017}; "Y" is assigned if the angular separation between the ALMA and Chandra sources is less than three times their combined $1\sigma$ positional error (see also \cite{ueda2018}).}}\\
\footnotemark[a]{The systematic coordinate offsets have been corrected.}\\
\footnotemark[b]{{The spectroscopic redshift presented by \citet{inami2017} using MUSE}}\\
\footnotemark[c]{{The photometric redshift presented by \citet{luo2017}.}}
\end{tabnote}
}
\end{sidewaystable*}

\subsection{Redshift distribution of ASAGAO sources} \label{subsec:redshift}

{\citet{straatman2016} estimate photometric redshifts of the ZFOURGE sources using the optical-to-near-IR SED fitting code \texttt{EAZY} \citep{brammer2008}. Their SED fitting is based on 40 photometric points from $U$-band to {\it Spitzer} 8-$\mu$m band including the \texttt{FourStar} 6 medium-band filters ($J_\mathrm{1},\ J_\mathrm{2},\ J_\mathrm{3},\ H_\mathrm{s},\ H_\mathrm{l},\ \mathrm{and}\ K_\mathrm{s}$-band, see Table 1 and Table 2 of \cite{straatman2016}, for details). Some ZFOURGE sources have spectroscopic redshifts presented by \citet{skelton2014}.} {One of the ASAGAO sources, ASAGAO ID26, has an extremely large photometric redshift ($z$ = 9.354), which is apparently caused by a incorrect SED fitting. On the other hand, \citet{luo2017} present its photometric redshift as $z$ = 2.14\footnote{{This value is obtained by the SED fitting of \citet{hsu2014}}} and this is the value we use. Some sources are also observed by \citet{inami2017} with the Multi Unit Spectroscopic Explorer (MUSE; \cite{bacon2010}). We use the spectroscopic redshifts of \citet{inami2017} for ASAGAO sources that are detected by MUSE.} 

As shown in Table \ref{tab:counterpart}, some ASAGAO sources have X-ray counterparts obtained by the {\it Chandra} deep field-south survey \citep{luo2017}. Therefore, some ASAGAO sources appear to have active galactic nuclei (AGNs). However, \citet{cowley2016} suggest that photometric redshifts estimated by ZFOURGE are appropriate for AGNs because of the benefits of medium-band filters. {We also have to note that \texttt{EAZY} adopts $K$-luminosity priors, but it does not affect our results significantly. We calculate absolute differences between estimated photometric redshifts with $K$-luminosity priors and without priors for ASAGAO sources without spectroscopic redshifts (i.e., 15 sources, see Table \ref{tab:counterpart}). The median value of the absolute differences is only 0.03.}

\begin{figure}[t!]
\begin{center}
\includegraphics[width=80mm]{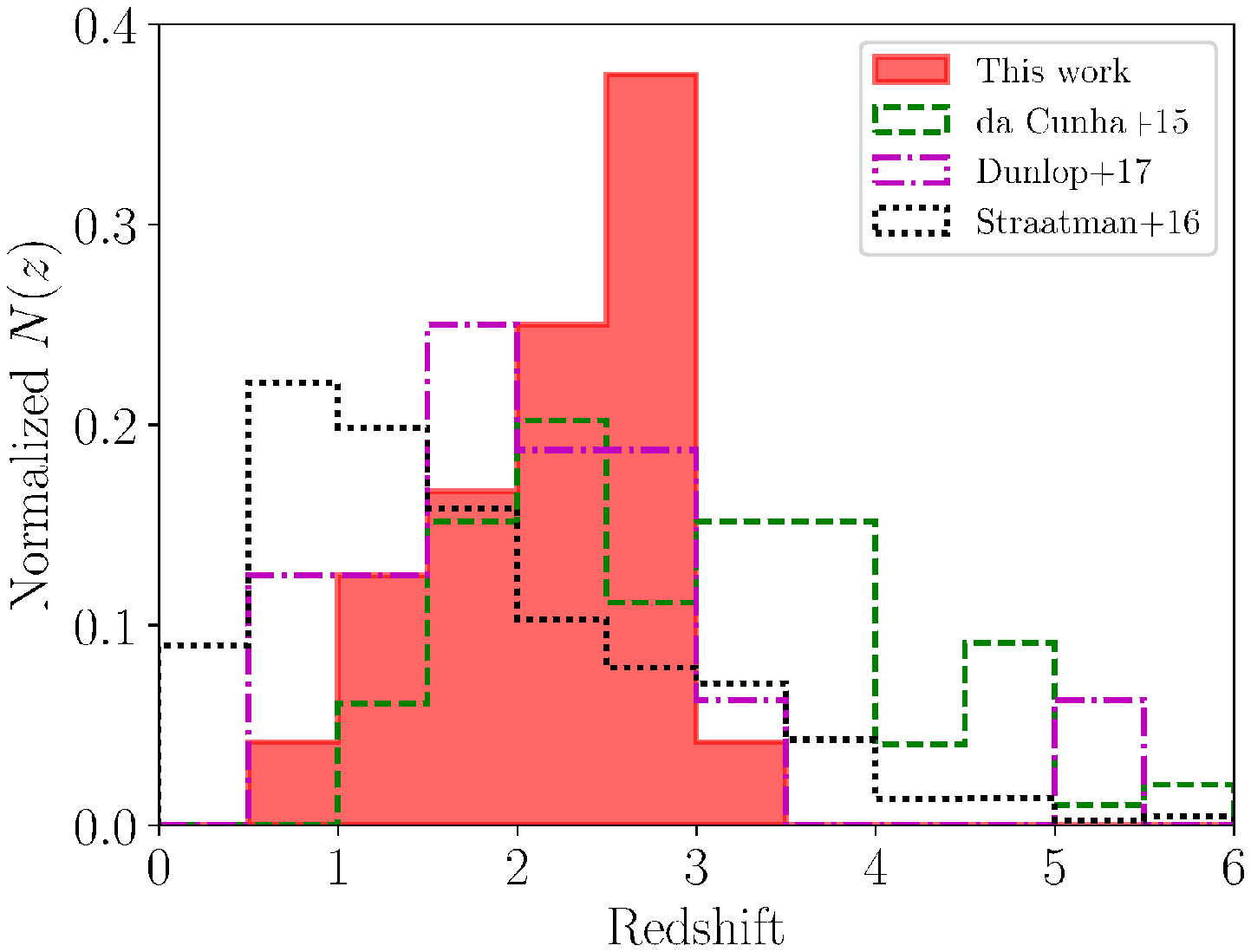}
\end{center}
\caption{Normalized redshift distribution of the {24} ASAGAO sources with ZFOURGE counterparts (red-shaded region). The green dashed line, magenta dot-dashed line, and black dotted line indicate redshift distribution of ALESS sources \citep{dacunha2015}, ALMA selected sources \citep{dunlop2017}, and ZFOURGE sources within ASAGAO field \citep{straatman2016}, respectively.
\label{fig:z_hist}}
\end{figure}

Figure \ref{fig:z_hist} shows the redshift distribution of ASAGAO sources. As a comparison, we also plot the results of ALESS \citep{dacunha2015}, ALMA detected sources with rest-frame optical/near-IR counterparts obtained by HUDF \citep{dunlop2017}, and ALMA non-detected ZFOURGE sources within the ASAGAO field \citep{straatman2016}. 
The median redshift of {24} ASAGAO sources is estimated to be {$z_\mathrm{median} = 2.39 \pm 0.14$}\footnote{{The median redshift of ALMA non-detected ZFOURGE sources is $z_\mathrm{median}$ = 1.45 $\pm$ 0.04}.}. This value is lower than that of ALESS sources ($z_\mathrm{median} = 2.83 \pm 0.22$; \cite{dacunha2015}), which are significantly brighter than ASAGAO sources, and rather similar to that in \citet{dunlop2017} sources, $z_\mathrm{median} = 2.04 \pm 0.29$ 
(although this is partly due to the fact that there are some overlaps between sources 
in ASAGAO and \citet{dunlop2017}). 

Many previous studies on ``classical'' SMGs ($S_\mathrm{1.2\ mm}\gtrsim$ a few mJy), including ALESS, report that median redshifts of ``classical'' SMGs are $z\sim$ 3, with a putative tail extending out to $z\sim6$ (e.g., \cite{chapman2005, simpson2014, dacunha2015, strandet2016}). On the other hand, \citet{aravena2016} suggest that their faint ALMA sources {with optical/near-IR counterparts} ($S_\mathrm{1.2\ mm}\sim$ 50--500 $\mu$Jy) reside in a lower redshift range than ``classical'' SMGs, although they only have small samples. {The similar trend between photometric redshifts and ALMA 870 $\mu$m flux density for SCUBA2-selected SMGs in UDS is also reported by \citet{stach2019}. They find a significant trend of increasing redshift with increasing 870 $\mu$m flux density, which exhibits a gradient of $dz/dS_{\rm 870\mu m}$ = 0.09 $\pm$ 0.02 mJy$^{-1}$ \citep{stach2019}.} {The redshift distribution of ASAGAO sources ($S_\mathrm{1.2\ mm}\lesssim$ 1 mJy)  is consistent with their results. Although we have to note that the difference of redshift distributions between (sub-)millimeter bright and faint sources can be caused by our sample selection (completenesses of optical/near-IR surveys drop significantly at high redshift), the difference is consistent with phenomenological models by \citet{betherman2015}, which suggest that the median redshift of (sub-)millimeter sources declines with decreasing flux densities. According to \citet{koprowski2017}, the fact that lower redshift sources tend to have lower (sub-)millimeter flux densities can be a direct consequence of the redshift evolution of the IR luminosity function (see also e.g., \citet{simpson2020})}.

\section{SED fitting from optical to millimeter wavelengths} \label{sec:SEDfit}

{In order to investigate the properties of dusty star-formation among ASAGAO sources, we have to estimate dust-obscured SFRs. Therefore, we compiled photometries from mid-IR to millimeter wavelengths to estimate IR luminosities accurately.} We include {\it Spitzer}/Multiband Imaging Photometer for the {\it Spitzer} (MIPS; \cite{rieke2004}, 24 $\mu$m), {\it Herschel}/Photodetector Array Camera and Spectrometer (PACS; \cite{poglitsch2010}, 100 and 160 $\mu$m), and {\it Herschel}/SPIRE (250, 350, and 500 $\mu$m) photometries, in addition to ZFOURGE data. {\it Spitzer}/MIPS 24 $\mu$m images are taken by \cite{dickinson2007} and its 1$\sigma$ is 3.9 $\mu$Jy \citep{straatman2016}. {\it Herschel}/PACS images are taken by \cite{magnelli2013} and their 1$\sigma$ are 205 and 354 $\mu$Jy at 100 and 160 $\mu$m, respectively \citep{straatman2016}. 

{For {\it Herschel}/SPIRE bands, we estimate de-blended flux densities
by adopting the de-blending technique that has been described in detail in \citet{liu2018}. 
Here
we have used all 24 $\mu$m and radio continuum sources as priors to extract fluxes in Herschel bands. From short to long wavelengths, after extracting source fluxes in shorter wavelength, we updated the flux prediction at longer wavelength. With this predicted flux, we updated the prior list for extraction at longer wavelength, for sources with predicted fluxes below the detection depth (typically 2-3 times the instrumental noise), we have frozen their fluxes to be the best predicted flux during the source extraction at longer wavelength, to reduce their effect on the source extraction for bright sources.  In the end, we only count extracted flux for sources that are not frozen as real measurements. We then run Monte Carlo simulations by injecting sources into real maps and re-do the source extraction together with true priors to estimate the accuracy for flux and flux uncertainties.
The typical flux uncertainties of de-blended SPIRE fluxes are estimated to be 2 to 3 mJy, which are similar to those in \citet{liu2018}.
The details of the de-blending procedure in the ASAGAO field will be presented in T.~Wang et al.~(in preparation).}

In this study, we perform bayesian-based SED fitting from optical to millimeter wavelengths using \texttt{MAGPHYS} (see \cite{dacunha2008,dacunha2015}, for details) to estimate the physical properties of the ASAGAO sources. {We adopt the SED templates of \citet{BC03} and the dust extinction model of \citet{charlot2000}.} In the SED fitting, we fixed the redshift of the ASAGAO sources to the best-fit photometric redshift presented by \cite{straatman2016} or spectroscopic redshift if available (see Table \ref{tab:counterpart}). Even if we consider the redshift uncertainties, our conclusions do not change significantly. For example, the changes in the estimated physical parameters are within $\lesssim$ 0.3 dex. {Although we consider photometry errors in each band, we do not consider systematic uncertainties (e.g., absolute flux calibration errors)\footnote{For example, according to ALMA Cycle 3 proposer's guide, the absolute flux calibration uncertainty of Band 6 data is expected to be $<$ 10\%.}, which does not affect our SED fitting results significantly.} For ASAGAO sources, we use the \texttt{MAGPHYS} high-$z$ extension version. {This code uses priors which are optimized for IR luminous dusty star-forming galaxies at high redshift \citep{dacunha2015}.} 

We have to note that \texttt{MAGPHYS} ignores any contribution by an AGN. Although \citet{hainline2011} suggest that the near-IR continuum excess can be caused by the AGNs, only 11\% of their sample ($\simeq$ 70 bright SMGs from \cite{chapman2005}) show stronger AGN-contribution than stellar-contribution at near-IR wavelengths. They also suggest that nearly half of their sample has less than 10\% AGN-contribution to the near-IR emissions (the median value seems to be $\sim$ 10--20\%, according to Figure 6 of \cite{hainline2011}). \citet{dunlop2017} suggest that an AGN component in faint (sub-)millimeter sources would contribute only $\simeq$ 20\% to the IR luminosity and near identical values are obtained by simply fitting the star-forming component to the ALMA data points. \citet{michaowski2014} also suggest that the contribution of the AGNs does not have any significant impact on the derived stellar masses of (sub-)millimeter sources, although some bright SMGs contain very luminous AGNs (e.g., \cite{ivison1998}) and the near-ubiquity of accreting black holes in SMGs are reported (e.g., \cite{alexander2005}). In the case of ASAGAO detected sources, \citet{ueda2018} suggest that majority of $X$-ray detected ASAGAO sources appear to be star-formation-dominant populations. 
Based on these considerations, in the following analysis we assume that the contribution from an AGN (if any) will have negligible impact on the physical properties derived from the SED analysis.

\begin{figure*}
\begin{center}
\includegraphics[scale = 0.68, angle = 90]{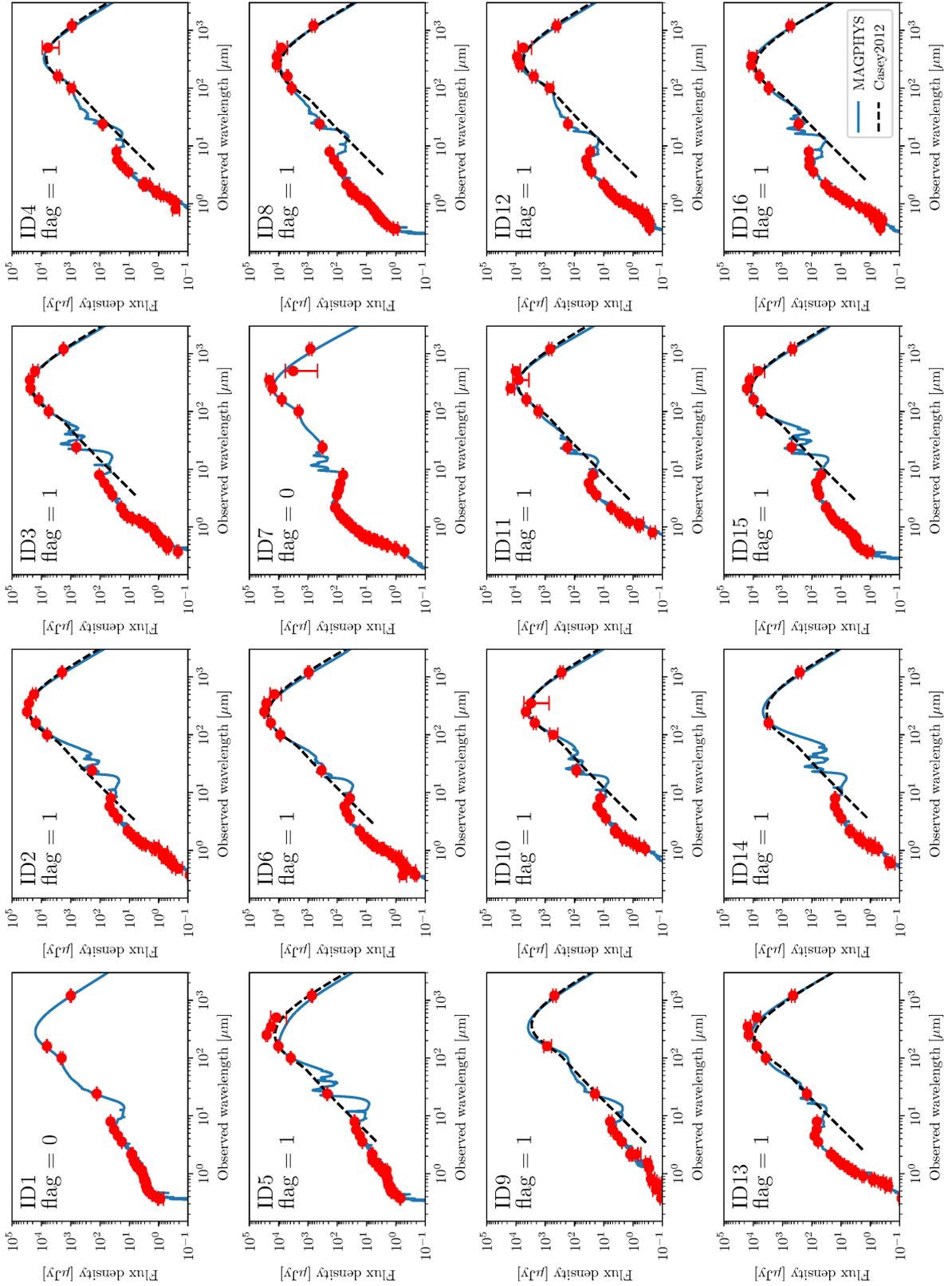}
\end{center}
\caption{Estimated SEDs of ASAGAO sources. Red symbols with errors are observed flux densities. Blue solid lines are the best fit SEDs estimated by \texttt{MAGPHYS} (see Section \ref{sec:SEDfit}). The black dashed lines are the best fit SEDs using a modified blackbody + mid-IR power-law model by \citet{casey2012}.
\label{fig:SED}}
\end{figure*}

\begin{figure*}
\begin{center}
\includegraphics[scale = 0.68, angle = 90]{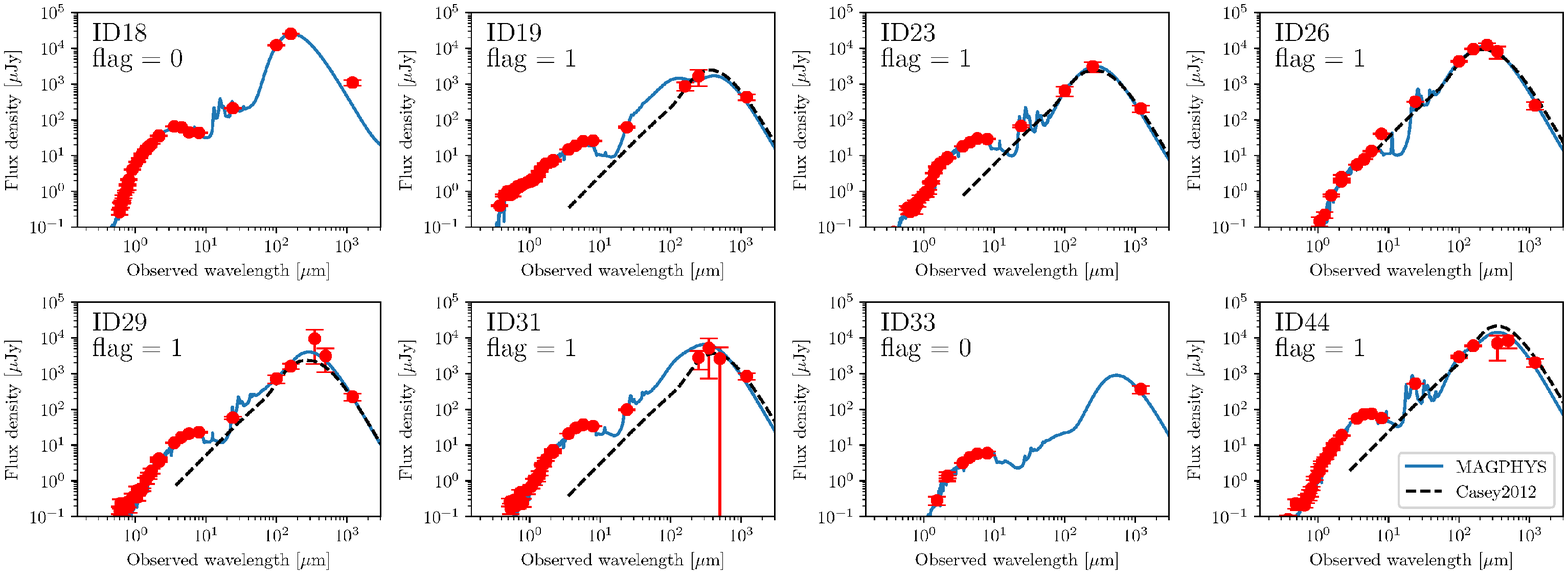}
\\Figure \ref{fig:SED} (Continued.)
\end{center}
\end{figure*}

The results of SED fitting are shown in Table \ref{tab:SEDresult} and Figure \ref{fig:SED}. In Table \ref{tab:SEDresult}, we add a flag to distinguish whether a source has a good (flag $=$ 1) or unreliable fit (flag $=$ 0). We manually remove 4 sources\footnote{
ID1 and ID7 because of large $p$-values, ID18 because of large discrepancy between its flux density at 1.2 mm and the best fit SED (Figure \ref{fig:SED}), ID33 because of the number of photometry points less than 12 (Figure \ref{fig:SED}). Note that ID18 may be affected by gravitational lensing by a chance coincidence of a foreground source. See also ID21 in Appendix 2.} with flag $=$ 0 from following discussion. 


\section{Physical properties} \label{sec:multi-w prop}

\subsection{Stellar masses and SFRs} \label{subsec:M-SFR}

We estimate stellar masses and SFRs of ASAGAO sources to discuss star formation properties. First, we calculate the stellar masses by using \texttt{MAGPHYS}. Second, we compute SFRs by summing the ultraviolet (UV) SFRs and IR SFRs based on the work of \citet{bell2005} scaled to a Chabrier IMF:
\begin{equation}
\mathrm{SFR_\mathrm{UV+IR}\ [M_\odot\ yr^{-1}]} = 1.09\times10^{-10}(L_\mathrm{IR}+2.2L_\mathrm{UV}).
\end{equation}
Here, $L_\mathrm{IR}\ [L_\odot]$ is the IR luminosity obtained by using \texttt{MAGPHYS}\footnote{Although \texttt{MAGPHYS} provides IR luminosities in the range of 3–-1000 $\mu$m in the rest-frame, the IR luminosities by \texttt{MAGPHYS} can be directly compared with other estimates referring to the commonly used wavelength range 8–-1000 $\mu$m in the rest-frame. This is because the contribution of dust to the emission in the range of 3–-8 $\mu$m is very small, as discussed in \citet{clemens2013}.}. The total UV luminosity, $L_\mathrm{UV}\ [L_\odot]$, is defined as $L_\mathrm{UV} = 1.5\nu L_{\nu2800}$ as described in \citet{straatman2016}, where $L_{\nu2800}$ is the rest-frame 2800 \AA\ luminosity. The derived stellar masses and SFRs are summarized in Table \ref{tab:SEDresult}.

\begin{figure}
\begin{center}
\includegraphics[width=80mm]{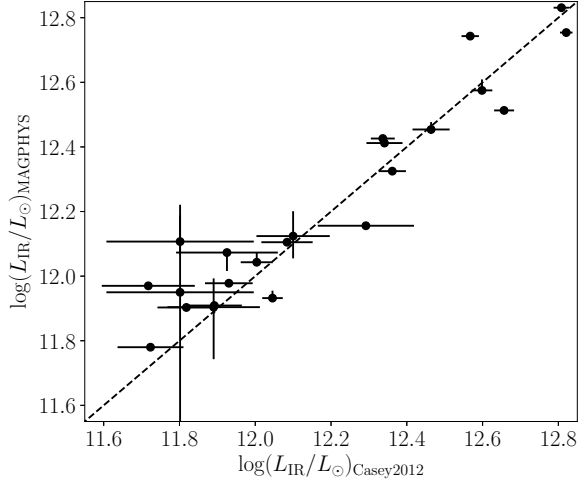}
\end{center}
\caption{Comparison between IR luminosities estimated using the SED model by \citet{casey2012} and \texttt{MAGPHYS}. The black dashed line indicates the case that $\log(L_\mathrm{IR}/L_\odot)_\mathrm{Casey2012}$ = $\log(L_\mathrm{IR}/L_\odot)_\mathrm{MAGPHYS}$.
\label{fig:L_IR}}
\end{figure}

We estimate the IR luminosities by mid-IR to far-IR SED templates obtained by \citet{casey2012} to confirm reliability of IR luminosities estimated by \texttt{MAGPHYS} for sources with flag $=$ 1 (Table \ref{tab:SEDresult}). \citet{casey2012} assume a modified blackbody radiation plus a mid-IR power law SED. Here, we assume an emissivity index equals 1.6 and mid-IR slope of 1.5 as discussed in \citet{casey2012}. As shown in Table \ref{tab:SEDresult} and Figure \ref{fig:L_IR}, there is no significant systematic offset between the two methods. 

\begin{figure*}
\begin{center}
\includegraphics[width=160mm]{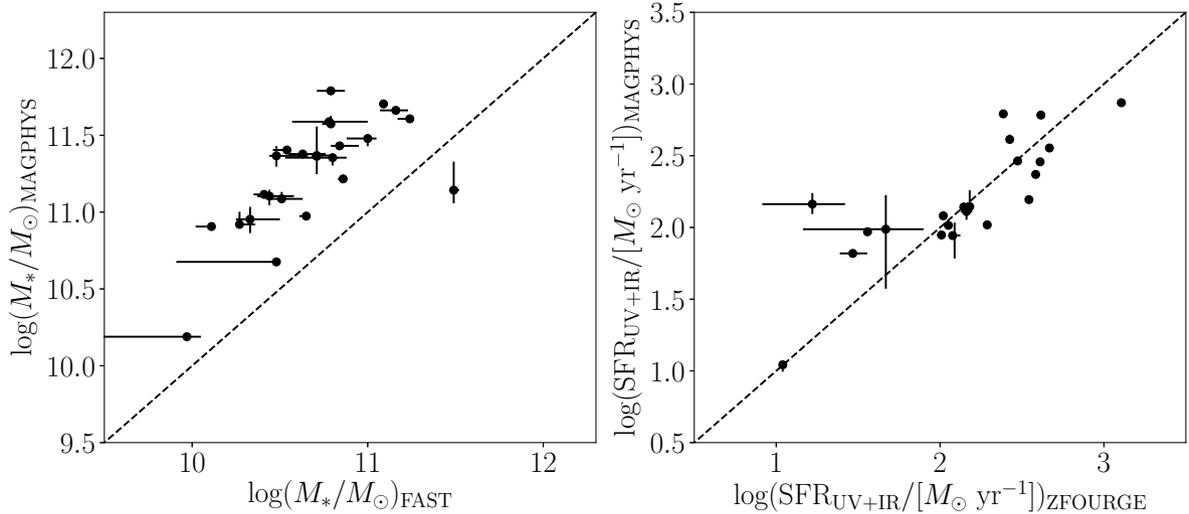}
\end{center}
\caption{(Left) Comparison of stellar masses obtained by \texttt{MAGPHYS} and \texttt{FAST}. The black dashed line indicates the case that $\log(M_*/M_\odot)_\mathrm{MAGPHYS}$ = $\log(M_*/M_\odot)_\mathrm{FAST}$.  (Right)  Comparison of IR + UV SFRs obtained by \texttt{MAGPHYS} and \texttt{ZFOURGE}. The black dashed line indicates the case that $\log(\mathrm{SFR_{UV+IR}})_\mathrm{MAGPHYS}$ = $\log(\mathrm{SFR_{UV+IR}})_\mathrm{ZFOURGE}$.
\label{fig:magphys_zfourge}}
\end{figure*}

In Table \ref{tab:SEDresult}, we show the stellar masses and SFRs of ASAGAO sources obtained by \citet{straatman2016}. They used the \texttt{FAST} code \citep{kriek2009} to derive stellar masses. For estimating UV+IR SFRs, they used IR luminosities obtained by the IR SED template of \citet{wuyts2008} in conjunction with MIPS 24 $\mu$m, PACS 100 $\mu$m, and PACS 160 $\mu$m photimetries and UV luminosities from the rest-frame 2800\ \AA\ luminosity. We compare our results with the ZFOURGE to check consistency in Figure \ref{fig:magphys_zfourge}. Although the SFRs estimated as with \texttt{MAGPHYS} and ZFOURGE are consistent, the stellar masses obtained by using \texttt{MAGPHYS} are systematically higher than that of \texttt{FAST} by $\gtrsim$ 0.2--0.5 dex. {A similar offset is also reported by \citep{michaowski2014} and they suggest that it can be explained by the difference of the assumed star formation histories. \citet{debarros2014} suggest that nebular emission lines at near-IR wavelengths, which are not included in \texttt{MAGPHYS}, can lead to an overestimation of the stellar masses.} Here, we use stellar masses obtained with \texttt{FAST} to compare our results with the ZFOURGE results (estimated by \texttt{FAST}) directly. {In this paper, we compare the derived stellar masses of ASAGAO sources with stellar masses of other (sub-)millimeter selected samples obtained by previous studies. Therefore, we have to note the differences of stellar mass modeling. For example, \citet{yamaguchi2016} also used \texttt{FAST} to estimate stellar masses. However, \citet{dacunha2015} used \texttt{MAGPHYS}, and \citet{dunlop2017} estimate stellar masses of ALMA sources by their SED fit using \citet{BC03} evolutionary synthesis models.}

\begin{figure*}
\begin{center}
\begin{tabular}{cc}
\begin{minipage}{0.45\hsize}
\begin{center}
\includegraphics[width = 75mm]{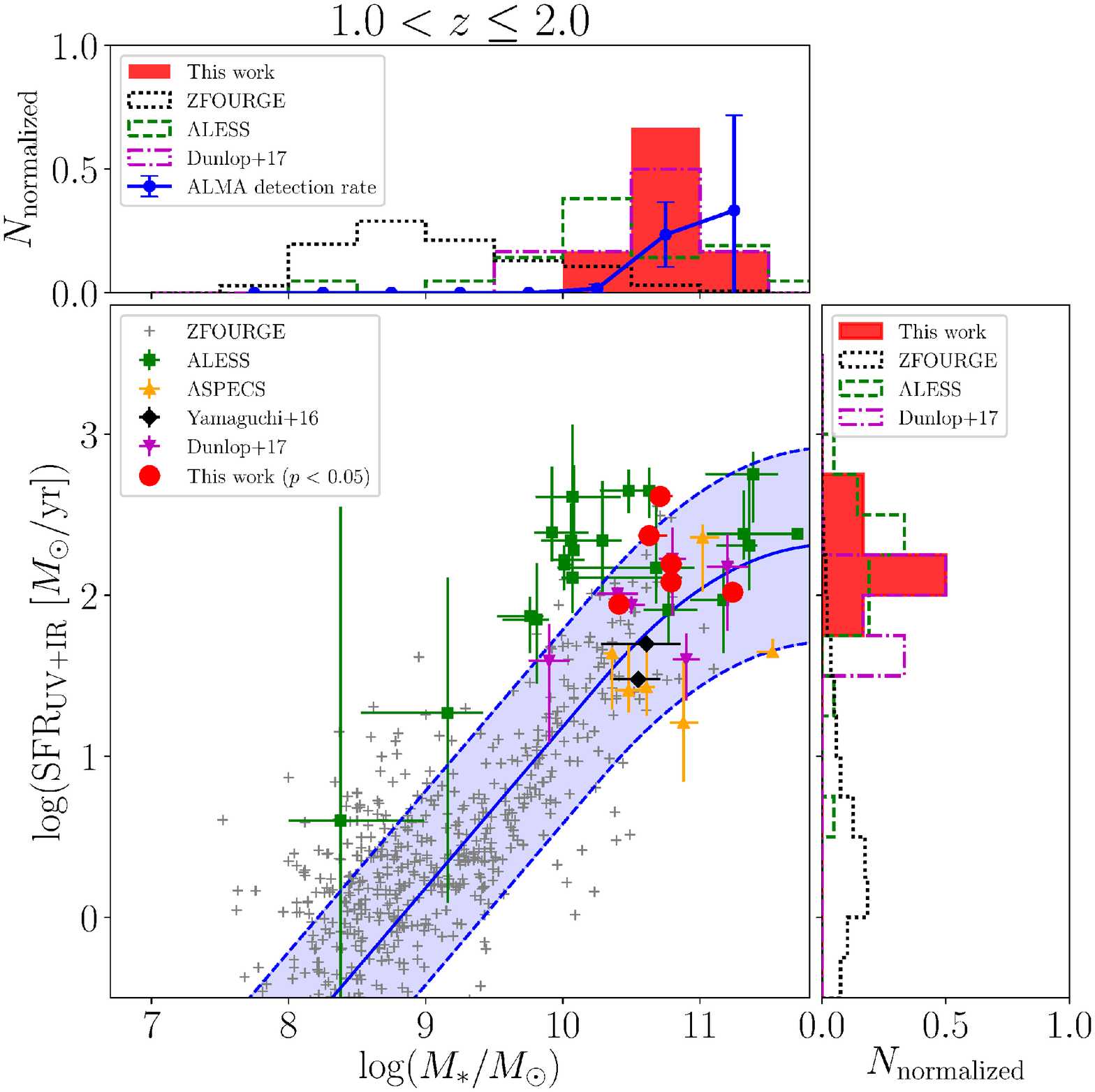}
\end{center}
\end{minipage}
\begin{minipage}{0.45\hsize}
\begin{center}
\includegraphics[width = 75mm]{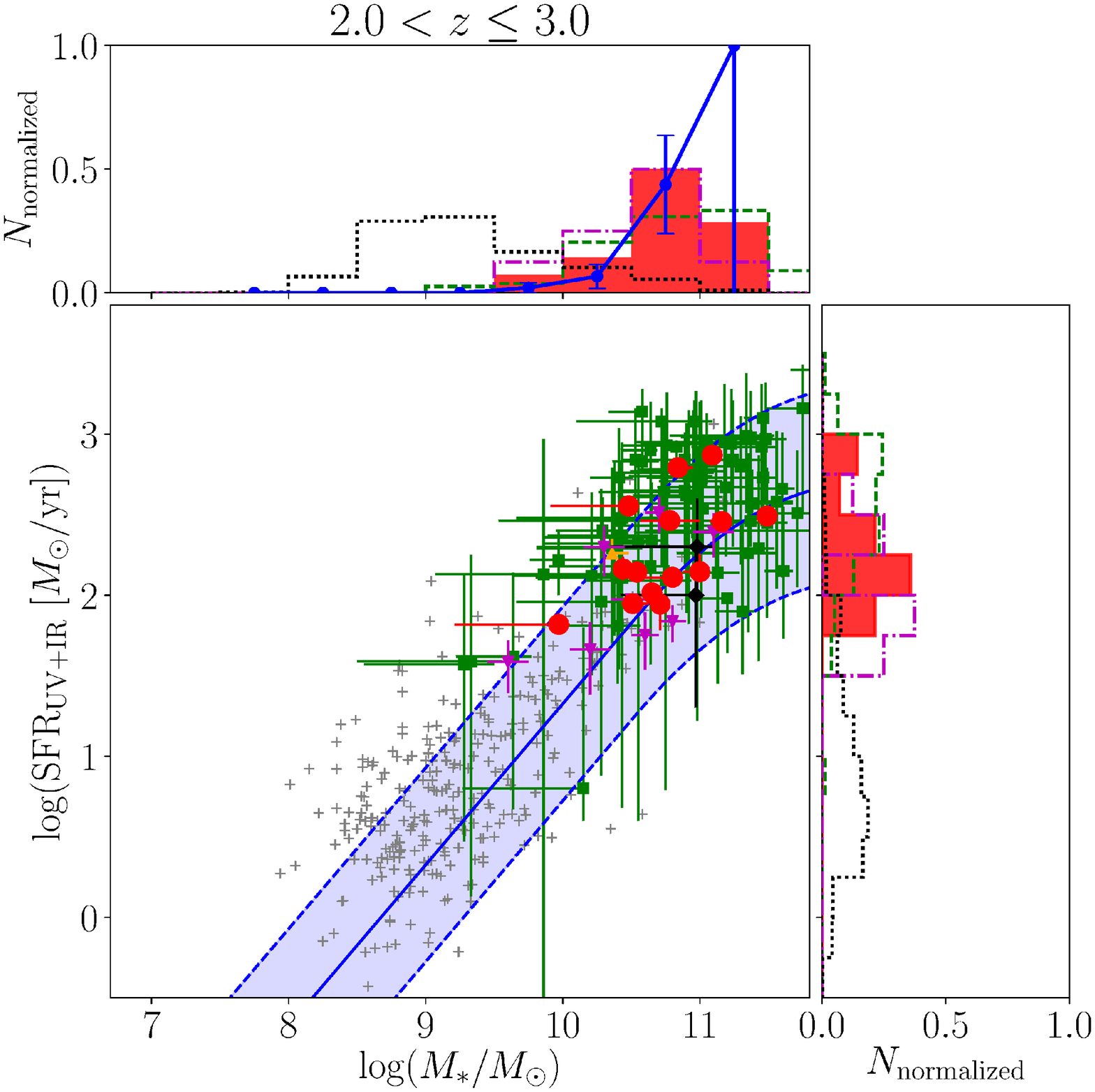}
\end{center}
\end{minipage}
\end{tabular}
\end{center}
\caption{Comparison of the stellar masses and SFRs of ASAGAO sources with the ``main sequence of star-forming galaxies''. ASAGAO sources are plotted as red circles. The gray crosses, green squares, orange triangles, black diamonds, and magenta inverse-triangles represent the ALMA non-detected ZFOURGE sources \citep{straatman2016}, ALESS sources \citep{dacunha2015}, ASPECS sources \citep{aravena2016}, faint SMGs of \cite{yamaguchi2016}, and ALMA selected sources by \citet{dunlop2017}. The blue solid lines indicate the position of the main sequence of star-forming galaxies at {$z = $ 1.83 (left) and 2.53 (right)} as predicted by \citet{schreiber2015}. The blue dashed lines indicate a factor of 4 above or below this main sequence. In addition, we show the histograms of stellar masses and SFRs. The blue circles in the stellar-mass distributions are ALMA detection rates as a function of their stellar masses. The error bars show simple Poisson uncertainties.
\label{fig:M_vs_SFR}}
\end{figure*}

Figure \ref{fig:M_vs_SFR} shows the stellar mass distribution of ASAGAO sources. We only include ASAGAO sources with SED fitting flag $=$ 1. Here, we divide ASAGAO sources into {two} redshift bins (i.e., $1.0<z\leq 2.0$, and $2.0<z\leq 3.0$). {We have to note that ID7 and ID33 are both excluded here even if they lie at $z<$ 1.0 and $z>$ 3.0, respectively.} In each redshift bin, there are 7 and 15 ASAGAO sources, respectively. The median stellar masses of each redshift bin are {$\log(M_*/M_\odot) =$ 10.75 $\pm$ 0.10 and 10.75 $\pm$ 0.11 for $1.0<z\leq 2.0$, and $2.0<z\leq 3.0$, respectively.} The estimated stellar masses are consistent with previous studies on ALMA continuum sources at similar redshift range and with $S_\mathrm{obs}\simeq$ 1 mJy such as \citet{tadaki2015} and \citet{dunlop2017}. As shown in Figure \ref{fig:M_vs_SFR}, the ASAGAO sources have typically higher stellar masses than ALMA\textcolor{blue}{-}non-detected ZFOURGE sources\footnote{{Herein we only use the star-forming galaxies selected by the {\it UVJ}-technique, as presented by \citet{whitaker2011}.}}, whose median stellar masses are {$\log(M_*/M_\odot) =$ 8.96 $\pm$ 0.05 and 9.17 $\pm$ 0.04 for $1.0<z\leq 2.0$, and $2.0<z\leq 3.0$, respectively.} This trend can be clearly observed when we plot the ALMA detection rate (i.e., ALMA-detected ZFOURGE sources per all ZFOURGE sources within the ASAGAO field) as a function of their stellar masses (Figure \ref{fig:M_vs_SFR}). {The trend is also shown in previous ALMA survey such as \citet{bouwens2016}.} Figure \ref{fig:M_vs_SFR} shows the SFR distribution of ASAGAO sources in two redshift bins. The median SFR of each redshift bin is {$\log(\mathrm{SFR}/[M_\odot\ \mathrm{yr}^{-1}])$ = 2.14 $\pm$ 0.13 and 2.15 $\pm$ 0.14 for $1.0<z\leq 2.0$ and $2.0<z\leq 3.0$, respectively.}

In Figure \ref{fig:M_vs_SFR}, we plot the ASAGAO sources on the $M_*$--SFR plane. In addition, we show the ALMA non-detected ZFOURGE sources within the ASAGAO field \citep{straatman2016}, ALESS sources \citep{dacunha2015}, ASPECS sources \citep{aravena2016}, faint SMGs in SXDF-UDS-CANDELS \citep{yamaguchi2016}, and ALMA sources {with optical/near-IR counterparts} by \citet{dunlop2017}. For comparison, we also plot the position of the main sequence of star-forming galaxies at each redshift {($z$ =1.83, and 2.53; median redshifts of each redshift bin)} compiled by \citet{schreiber2015}.

As shown in Figure \ref{fig:M_vs_SFR}, the ASAGAO sources primarily lie on the main sequence of star-forming galaxies, although some ASAGAO sources shows starburst-like features. 
{Here we adopt the definition of a ``starburst'' mode by \citet{schreiber2015}, where an SFR increased by more than a factor 4 (or 0.6 dex) compared to the main sequence.}
This is consistent with previous ALMA results (e.g., \cite{dacunha2015, aravena2016, yamaguchi2016, dunlop2017}). Figure \ref{fig:M_vs_SFR} also suggests that ASAGAO sources mainly trace the high-mass end of the main sequence of star-forming galaxies. When we compare ASAGAO sources with ALESS sources (i.e., single-dish selected galaxies), ASAGAO sources tend to have systematically lower SFRs for a similar stellar mass range. Here we need to note that \citet{dacunha2015} used \texttt{MAGPHYS} to estimate stellar masses of ALESS sources. When we consider the systematic offset of stellar masses estimated by \texttt{MAGPHYS} and \texttt{FAST}, differences between ASAGAO sources and ALESS sources on the $M_*$-SFR plane become even larger. This result implies that an ALMA continuum survey at a 1$\sigma$ depth of a few tens of $\mu$Jy can unveil galaxies which are more likely the normal star-forming galaxies than ``classical'' SMGs {since they show more quiescent star-forming activities than ``classical'' SMGs for a similar stellar mass range}. 

\begin{sidewaystable*}[]
\small{
\caption{Results of the SED fitting \label{tab:SEDresult}}
\begin{center}
\begin{tabular}{ccccccccccc}
\hline\hline
{ID} &
{ID} & {$\log(M_\mathrm{*})$}  & {$\log(M_\mathrm{*})$} & {$\log(L_\mathrm{IR})$} & {$\log(L_\mathrm{IR})$} & 
{$\log(L_\mathrm{UV})$} & {$\log(\mathrm{SFR_{UV+IR}})$} & {$\log(\mathrm{SFR_{UV+IR}})$} & {$\beta_\mathrm{UV}$} & {flag} \\
{(ZFOURGE)} & {(ASAGAO)} & {(ZFOURGE)} & {(\texttt{MAGPHYS})} & {(\texttt{MAGPHYS})} & {(\cite{casey2012})} & 
{(\texttt{MAGPHYS})} & {(ZFOURGE)}  & {(\texttt{MAGPHYS})} & {} & {} \\
{} & {} &{$[M_\odot]$} & {$[M_\odot]$} & {$[L_\odot]$} & {$[L_\odot]$} & 
{$[L_\odot]$} & {$[M_\odot\ \mathrm{yr}^{-1}]$}  & {$[M_\odot\ \mathrm{yr}^{-1}]$} & {} & {} \\
{(1)} & {(2)} & {(3)} & {(4)} & {(5)} & {(6)} & {(7)} & {(8)} & {(9)} & {(10)} & \textcolor{blue}{(11)} \\
\hline
18658 & 1 & 10.11$^{+0.00}_{-0.09}$ & 10.91$^{+0.00}_{-0.00}$ & 12.74$^{+0.00}_{-0.00}$ & -- & 10.26$^{+0.04}_{-0.04}$ & 2.62$^{+0.01}_{-0.01}$ & 2.78$^{+0.00}_{-0.00}$ & -1.53$\pm$0.07 & 0\\
17856 & 2 & 10.84$^{+0.11}_{-0.05}$ & 11.43$^{+0.00}_{-0.00}$ & 12.75$^{+0.00}_{-0.00}$ & 12.82$\pm$0.02 & 9.63$^{+0.06}_{-0.07}$ & 2.39$^{+0.01}_{-0.01}$ & 2.79$^{+0.00}_{-0.00}$ & 0.14$\pm$0.27 & 1\\
13086 & 3 & 11.09$^{+0.00}_{-0.01}$ & 11.70$^{+0.00}_{-0.00}$ & 12.83$^{+0.00}_{-0.00}$ & 12.81$\pm$0.02 & 9.97$^{+0.05}_{-0.06}$ & 3.11$^{+0.00}_{-0.00}$ & 2.87$^{+0.00}_{-0.00}$ & -0.43$\pm$0.16 & 1\\
18645 & 4 & 10.78$^{+0.22}_{-0.21}$ & 11.59$^{+0.00}_{-0.01}$ & 12.43$^{+0.00}_{-0.00}$ & 12.34$\pm$0.03 & 8.94$^{+0.19}_{-0.36}$ & 2.47$^{+0.02}_{-0.02}$ & 2.46$^{+0.00}_{-0.00}$ & -- & 1\\
18701 & 5 & 10.48$^{+0.00}_{-0.57}$ & 10.68$^{+0.00}_{-0.00}$ & 12.51$^{+0.00}_{-0.00}$ & 12.66$\pm$0.03 & 10.08$^{+0.03}_{-0.04}$ & 2.67$^{+0.01}_{-0.01}$ & 2.55$^{+0.00}_{-0.00}$ & -1.45$\pm$0.14 & 1\\
22177 & 6 & 10.71$^{+0.09}_{-0.03}$ & 11.37$^{+0.19}_{-0.03}$ & 12.57$^{+0.03}_{-0.00}$ & 12.60$\pm$0.03 & 9.69$^{+0.06}_{-0.07}$ & 2.43$^{+0.01}_{-0.01}$ & 2.61$^{+0.03}_{-0.00}$ & -0.51$\pm$0.18 & 1\\
20298 & 7 & 10.27$^{+0.09}_{-0.02}$ & 10.92$^{+0.08}_{-0.00}$ & 11.00$^{+0.00}_{-0.04}$ & -- & 8.87$^{+0.22}_{-0.47}$ & 1.04$^{+0.02}_{-0.01}$ & 1.04$^{+0.01}_{-0.05}$ & -- & 0\\
19033 & 8 & 11.16$^{+0.07}_{-0.09}$ & 11.66$^{+0.00}_{-0.00}$ & 12.41$^{+0.00}_{-0.00}$ & 12.34$\pm$0.05 & 10.36$^{+0.05}_{-0.06}$ & 2.61$^{+0.01}_{-0.00}$ & 2.46$^{+0.00}_{-0.00}$ & -0.43$\pm$0.09 & 1\\
21234 & 9 & 9.97$^{+0.08}_{-0.76}$ & 10.19$^{+0.00}_{-0.00}$ & 11.78$^{+0.00}_{-0.00}$ & 11.72$\pm$0.09 & 9.03$^{+0.13}_{-0.20}$ & 1.47$^{+0.09}_{-0.08}$ & 1.82$^{+0.00}_{-0.00}$ & -0.61$\pm$0.80 & 1\\
18912 & 10 & 10.51$^{+0.12}_{-0.08}$ & 11.09$^{+0.04}_{-0.02}$ & 11.91$^{+0.01}_{-0.02}$ & 11.89$\pm$0.07 & 8.94$^{+0.17}_{-0.27}$ & 2.01$^{+0.03}_{-0.02}$ & 1.95$^{+0.01}_{-0.02}$ & -- & 1\\
21730 & 11 & 10.54$^{+0.01}_{-0.08}$ & 11.40$^{+0.00}_{-0.00}$ & 12.11$^{+0.00}_{-0.00}$ & 12.08$\pm$0.07 & 8.77$^{+0.27}_{-0.85}$ & 2.14$^{+0.02}_{-0.01}$ & 2.14$^{+0.00}_{-0.00}$ & -- & 1\\
16952 & 12 & 10.41$^{+0.05}_{-0.06}$ & 11.12$^{+0.00}_{-0.00}$ & 11.90$^{+0.00}_{-0.00}$ & 11.82$\pm$0.08 & 9.42$^{+0.08}_{-0.10}$ & 2.08$^{+0.02}_{-0.01}$ & 1.94$^{+0.00}_{-0.00}$ & -0.59$\pm$0.37 & 1\\
17733 & 13 & 10.79$^{+0.00}_{-0.05}$ & 11.57$^{+0.05}_{-0.00}$ & 12.04$^{+0.03}_{-0.00}$ & 12.00$\pm$0.04 & 9.08$^{+0.23}_{-0.50}$ & 2.02$^{+0.02}_{-0.01}$ & 2.08$^{+0.03}_{-0.00}$ & -2.03$\pm$0.02 & 1\\
18336 & 14 & 10.44$^{+0.14}_{-0.07}$ & 11.10$^{+0.04}_{-0.06}$ & 12.12$^{+0.08}_{-0.07}$ & 12.10$\pm$0.10 & 9.27$^{+0.09}_{-0.11}$ & 1.22$^{+0.20}_{-0.31}$ & 2.16$^{+0.08}_{-0.07}$ & -- & 1\\
15702 & 15 & 10.63$^{+0.13}_{-0.08}$ & 11.38$^{+0.00}_{-0.00}$ & 12.32$^{+0.00}_{-0.00}$ & 12.36$\pm$0.04 & 10.21$^{+0.06}_{-0.07}$ & 2.58$^{+0.01}_{-0.00}$ & 2.37$^{+0.00}_{-0.00}$ & -1.38$\pm$0.19 & 1\\
19487 & 16 & 11.24$^{+0.00}_{-0.07}$ & 11.61$^{+0.00}_{-0.00}$ & 11.98$^{+0.00}_{-0.00}$ & 11.93$\pm$0.06 & 9.45$^{+0.18}_{-0.30}$ & 2.29$^{+0.01}_{-0.01}$ & 2.02$^{+0.00}_{-0.00}$ & -0.05$\pm$0.61 & 1\\
12438 & 18 & 10.48$^{+0.18}_{-0.04}$ & 11.37$^{+0.06}_{-0.07}$ & 11.93$^{+0.02}_{-0.01}$ & -- & 8.67$^{+0.15}_{-0.23}$ & 1.56$^{+0.02}_{-0.01}$ & 1.97$^{+0.02}_{-0.01}$ & -- & 0\\
14580 & 19 & 10.65$^{+0.02}_{-0.04}$ & 10.97$^{+0.00}_{-0.00}$ & 11.97$^{+0.00}_{-0.00}$ & 11.72$\pm$0.12 & 9.87$^{+0.04}_{-0.05}$ & 2.05$^{+0.03}_{-0.02}$ & 2.02$^{+0.00}_{-0.00}$ & -0.69$\pm$0.16 & 1\\
18270 & 23 & 10.71$^{+0.03}_{-0.00}$ & 11.36$^{+0.02}_{-0.12}$ & 11.90$^{+0.09}_{-0.16}$ & 11.89$\pm$0.12 & 9.38$^{+0.08}_{-0.10}$ & 2.09$^{+0.04}_{-0.03}$ & 1.94$^{+0.09}_{-0.16}$ & -0.64$\pm$0.60 & 1\\
14146 & 26 & 11.49$^{+0.03}_{-0.00}$ & 11.14$^{+0.18}_{-0.09}$ & 12.45$^{+0.02}_{-0.01}$ & 12.46$\pm$0.05 & 8.14$^{+0.44}_{-0.11}$ & -- & 2.49$^{+0.02}_{-0.01}$ & -- & 1\\
14419 & 29 & 10.80$^{+0.08}_{-0.27}$ & 11.35$^{+0.00}_{-0.05}$ & 12.07$^{+0.00}_{-0.06}$ & 11.93$\pm$0.13 & 8.92$^{+0.29}_{-1.55}$ & 2.16$^{+0.03}_{-0.03}$ & 2.11$^{+0.00}_{-0.06}$ & -0.41$\pm$0.65 & 1\\
13714 & 31 & 11.00$^{+0.05}_{-0.12}$ & 11.48$^{+0.02}_{-0.05}$ & 12.11$^{+0.11}_{-0.04}$ & 11.80$\pm$0.19 & 9.12$^{+0.24}_{-0.59}$ & 2.18$^{+0.02}_{-0.02}$ & 2.15$^{+0.11}_{-0.04}$ & -1.07$\pm$1.04 & 1\\
14122 & 33 & 10.33$^{+0.17}_{-0.08}$ & 10.95$^{+0.08}_{-0.09}$ & 11.95$^{+0.24}_{-0.41}$ & -- & 8.49$^{+0.10}_{-0.13}$ & 1.67$^{+0.23}_{-0.51}$ & 1.99$^{+0.24}_{-0.41}$ & -- & 0\\
14700 & 44 & 10.79$^{+0.08}_{-0.08}$ & 11.79$^{+0.00}_{-0.00}$ & 12.16$^{+0.00}_{-0.00}$ & 12.29$\pm$0.13 & 9.04$^{+0.27}_{-0.82}$ & 2.54$^{+0.01}_{-0.00}$ & 2.19$^{+0.00}_{-0.00}$ & -1.94$\pm$0.76 & 1\\
\hline
\end{tabular}
\end{center}
\begin{tabnote}
{\bf Notes.} {(1) ZFOURGE ID (2) ASAGAO ID. (3) Stellar mass taken from the ZFOURGE catalog \citep{straatman2016}, which are obtained using \texttt{FAST} (4) Stellar mass obtained by \texttt{MAGPHYS}. (5) IR luminosity obtained by \texttt{MAGPHYS}. {(6) IR luminosity obtained using \citet{casey2012} model.} (7) UV luminosity obtained by rest-frame 2800\ \AA\ luminosity. (8) UV + IR SFR obtained by ZFOURGE \citep{straatman2016}. (9) UV + IR SFR obtained by \texttt{MAGPHYS}. (\textcolor{blue}{10}) UV spectral slope estimated by fitting a power law $f_\lambda \propto \lambda^{\beta}$ over the rest-frame wavelength range of 1500--2500 \AA. (11) SED fitting flag (1: good, 0: bad). There are the reasons to classify as flag $=$ 0: a) the number of photometry points is less than 12 (i.e., the degree of freedom of SED fit using \texttt{MAGPHYS} is less than one), b) the predicted millimeter photometry is inconsistent with the observed ALMA photometry and there are no photometry points at mid-IR-to-far-IR wavelengths. {c) The $p$-value is larger than 0.05.}}
\end{tabnote}
}
\end{sidewaystable*}

\subsection{The infrared excess (IRX)} \label{subsec:IRX}

As shown in Figure \ref{fig:M_vs_SFR}, there are ALMA\textcolor{blue}{-}non-detected ZFOURGE sources within the ASAGAO field even though they show similar star-forming properties to ALMA\textcolor{blue}{-}detected sources on the $M_*$--SFR plane. In this section, we focus on IRX (i.e., $L_\mathrm{IR}/L_\mathrm{UV}$) as a key parameter to distinguish between ALMA\textcolor{blue}{-}detected sources and non-detected sources. Although many previous studies on IRX of galaxies use rest-frame 1600 \AA\ luminosities, we note that we adopt $L_\mathrm{UV} = 1.5\nu L_{\nu2800}$ to obtain $L_\mathrm{UV}$ (see Section \ref{subsec:M-SFR}), which are supposed to be approximately equivalent \citep{kennicutt1998, whitaker2014}.

\begin{figure}
\begin{center}
\includegraphics[width=80mm]{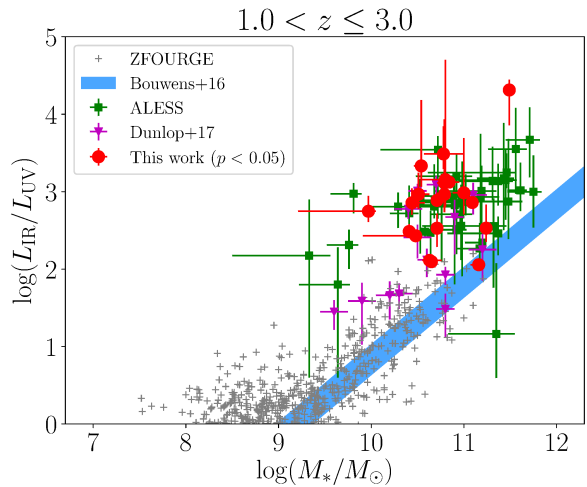}
\end{center}
\caption{The IRX of ASAGAO sources as a function of their stellar mass (red circles). We also show the ALMA\textcolor{blue}{-}non-detected ZFOURGE sources \citep{straatman2016} within the ASAGAO field, ALESS sources \citep{dacunha2015}, and ALMA\textcolor{blue}{-}selected sources by \cite{dunlop2017}. The thick shaded blue line shows the {consensus relation compiled} by UV-selected galaxies at $z\sim$ 2--3 \citep{bouwens2016}.
\label{fig:M_vs_IRX}}
\end{figure}

\subsubsection{The IRX--$M_*$ and IRX--SFR relations} \label{subsubsec:IRX-M}

Several studies have shown a correlation between the IRX and stellar masses, in the sense that massive star-forming galaxies show larger IRX (e.g., \cite{reddy2010, whitaker2014, bouwens2016, dunlop2017}). We plot the IRX of ASAGAO sources as a function of their stellar masses in Figure \ref{fig:M_vs_IRX}. For comparison, we also show the ALMA\textcolor{blue}{-}detected sources \citep{dacunha2015, dunlop2017} and ALMA\textcolor{blue}{-}non-detected ZFOURGE sources \citep{straatman2016} within the ASAGAO field. {We also show the consensus IRX-$M_*$ relation compiled by \citet{bouwens2016}. They derive stellar masses using \texttt{FAST} and their estimated consensus relationship is consistent with the results of three separate studies \citep{reddy2010, whitaker2014, alvarez2016}.}

\begin{figure*}
\begin{center}
\includegraphics[width=160mm]{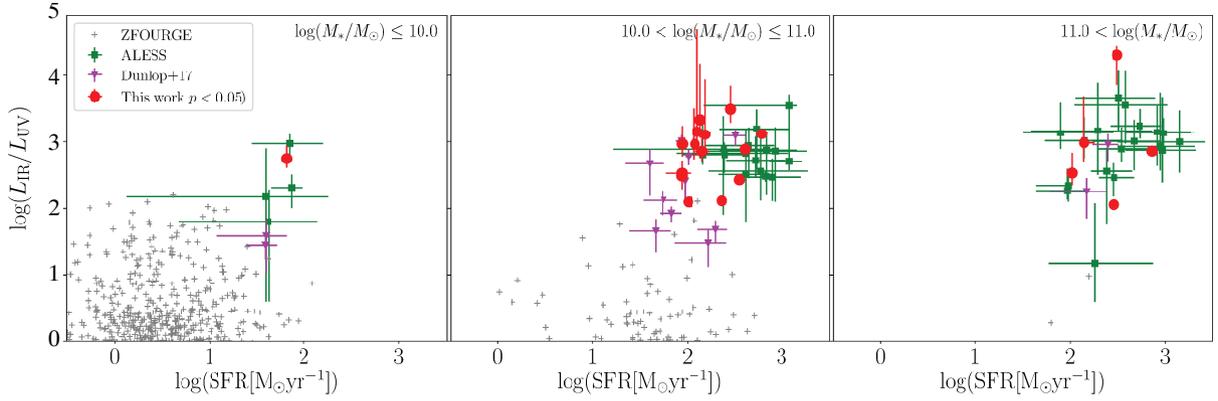}
\end{center}
\caption{The IRX of ASAGAO sources as a function of their SFRs (red circles). We also show the ALMA\textcolor{blue}{-}non-detected ZFOURGE sources within the ASAGAO field \citep{straatman2016}, ALESS sources \citep{dacunha2015}, and ALMA\textcolor{blue}{-}selected sources by \citet{dunlop2017}. 
\label{fig:SFR_vs_IRX}}
\end{figure*}

As shown in Figure \ref{fig:M_vs_IRX}, the ALMA\textcolor{blue}{-}detected sources tend to have larger IRX compared to the ALMA\textcolor{blue}{-}non-detected sources. The IRXs of ASAGAO sources at $z > 1.0$ are systematically larger than those from the IRX-$M_*$ relation of UV-selected galaxies, with an offset of 1--2 dex; in contrast, no ALMA\textcolor{blue}{-}non-detected ZFOURGE sources exhibit such elevated IRX values. {When we plot the IRX-SFR relation of ASAGAO sources for three stellar mass bins (i.e., $\log(M_*/M_\odot) \leq 10$, $10 < \log(M_*/M_\odot) \leq 11$, and $11< \log(M_*/M_\odot)$; Figure \ref{fig:SFR_vs_IRX}), the offset from ALMA\textcolor{blue}{-}non-detected ZFOURGE sources also become evident.}

%

\subsubsection{The IRX--$\beta_\mathrm{UV}$ relation} \label{subsubsec:IRX-beta}

\begin{figure*}
\begin{center}
\includegraphics[width=160mm]{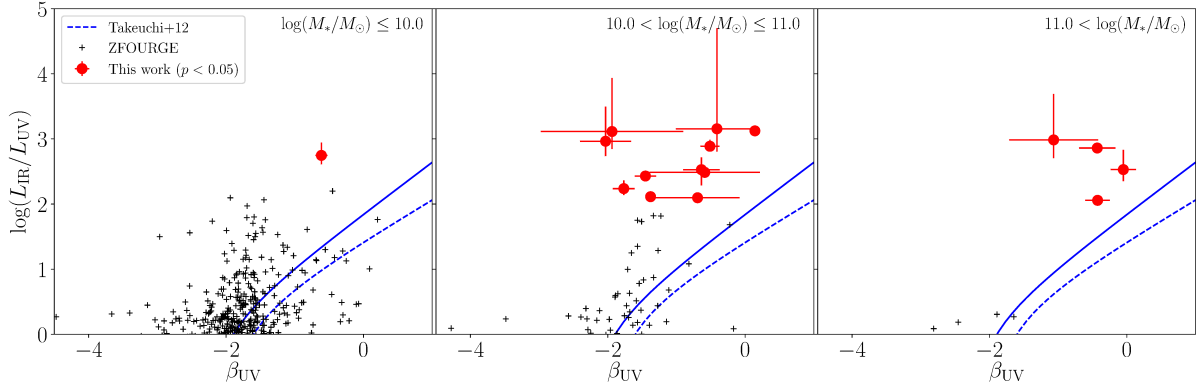}
\end{center}
\caption{The IRX of the ASAGAO sources as a function of $\beta_\mathrm{UV}$ (red circles). Black crosses indicate ALMA\textcolor{blue}{-}non-detected ZFOURGE sources \citep{straatman2016} within the ASAGAO field. {The blue solid and dashed lines are the IRX-$\beta_\mathrm{UV}$ relations of \citet{meuer1999} and \citet{takeuchi2012}, respectively.}
\label{fig:IRX-beta}}
\end{figure*}

An useful relation to study the properties of dust is the relation between the UV spectral slopes ($\beta_\mathrm{UV}$) and IRX, because this relation reflect the effect of dust attenuation. Therefore, we examine the IRX-$\beta_\mathrm{UV}$ relation of ALMA\textcolor{blue}{-}detected sources for further discussion on the difference between ALMA\textcolor{blue}{-}detected and non-detected sources. The IRX-$\beta_\mathrm{UV}$ relation has been calibrated using local star-burst galaxies (e.g., \cite{meuer1999, takeuchi2012}). 

In this study, $\beta_\mathrm{UV}$ is calculated by fitting a power law $f_\lambda \propto \lambda^{\beta}$ over the rest-frame wavelength range of 1500--2500 \AA\ using ZFOURGE photometies. Figure \ref{fig:IRX-beta} shows the IRX--$\beta_\mathrm{UV}$ relation of ASAGAO sources. We also plot the ALMA\textcolor{blue}{-}non-detected ZFOURGE sources within the ASAGAO field, along with the relation given in \citet{meuer1999} and \citet{takeuchi2012}. 
We find that ASAGAO sources tend to have larger IRX values compared to the ALMA\textcolor{blue}{-}non-detected ZFOURGE sources, as well as the local starburst relations as provided by \citet{meuer1999} and \citet{takeuchi2012}. This trend is consistent with the results in the COSMOS field by \citet{casey2014} and a recent update by \citet{fudamoto2020}, although some ASAGAO sources exhibit more elevated IRX values.


Then why do dusty star-forming galaxies lie above the local IRX-$\beta_\mathrm{UV}$ relations by \citet{meuer1999} and \citet{takeuchi2012}? 
One of the possible drivers is the difference in a starburst time-scale \citep{casey2014}. Dusty star-forming galaxies have short-timescale starburst (10--300 Myr), and short-lived burst events produce many young O and B stars that are not entirely enshrouded in thick dust cocoons yet.
Another driver for the elevated IRX values is the dust composition (i.e., the difference of the chemical composition or/and the grain size distribution; e.g., \cite{safarzadeh2017,galliano2018}).
The geometry of dust and stellar components will also have significant impact on the IRX-$\beta_{\rm UV}$ relation. 
In fact, starburst galaxies hosting heavily obscured regions together with a small fraction of non-obscured regions 
(e.g., ``holes in dust shields'') can easily deviate from the local relation, 
because their UV and IR fluxes 
no longer come from the same region of a galaxy 
(e.g., \cite{popping2017, narayanan2018, fudamoto2020}). 
Significant difference between the dust-obscured star-forming regions and less-obscured rest-UV-emitting regions 
has been reported by recent ALMA observations 
(e.g., \cite{hodge2015, tadaki2017, chen2017}), 
and ASAGAO sources discussed here are also reported to exhibit such difference 
\citep{fujimoto2018}.
Detailed comparison with the (sub)millieter and rest-UV distributions of these sources with higher resolution observations 
will be useful to quantitatively address the impact of dust-stellar geometry on the measured IRX-$\beta_{\rm UV}$ relations.

\medskip

It has also been claimed that dust temperature (e.g., \cite{faisst2017,narayanan2018}) and the presence of a low-level AGN (\cite{saturni2018}) can also affect IRX-$\beta_{UV}$ relations of duty sources. Spatially resolved, shorter-wave ALMA observations will be necessary to disentangle the impact of warm dust in these dusty galaxies.


\section{Contribution to the cosmic SFRD} \label{sec:SFRD}
In this section, we use our ASAGAO results to explore the evolution of the cosmic SFRD. Because of the high sensitivity and high angular resolution of ALMA, we can resolve the contribution of dusty star-forming sources to the cosmic SFRD down to $\log(L_\mathrm{IR}/L_\odot)\sim$ 11, which is $\sim$ 0.5--1 dex lower luminosity range than previous \textit{Herschel} observations at $z\gtrsim$ 2 (e.g., \cite{gruppioni2013}).

\begin{figure*}[t!]
\begin{center}
\includegraphics[width=160mm]{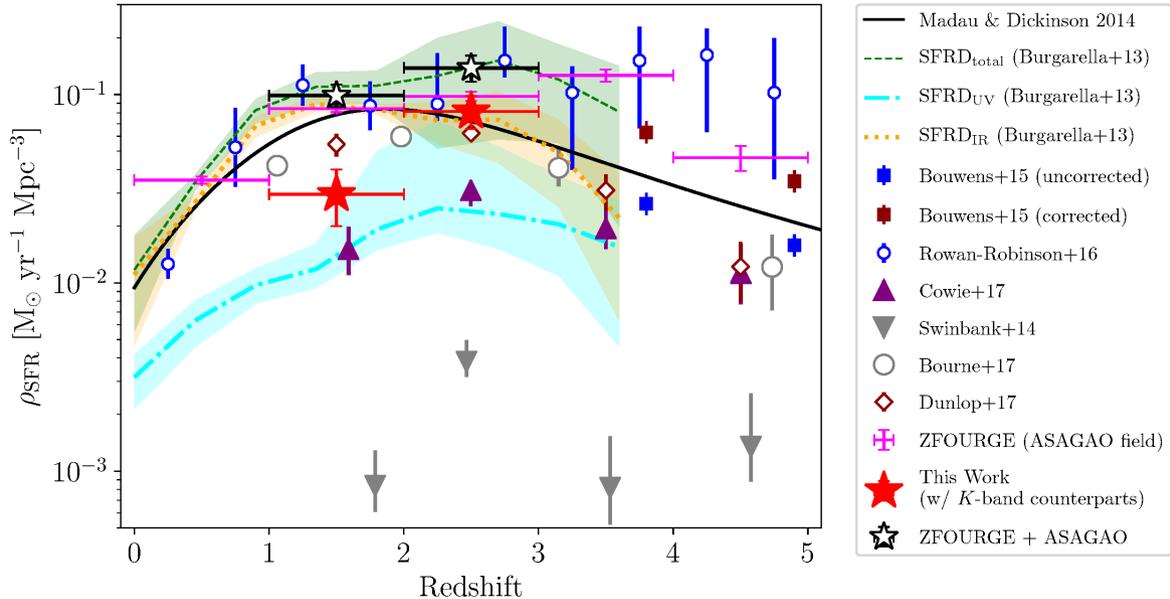}
\end{center}
\caption{{Contribution of ASAGAO sources to the cosmic SFRD as a function of redshift. Red, magenta, and black symbols are the contributions of ASAGAO sources with $K$-band counterparts, ALMA non-detected ZFOURGE sources within ASAGAO field, and their sum, respectively.} We adopt simple Poisson errors and {SFR-errors attributed to redshift uncertainties}. The black solid line indicate the recent results of the redshift evolution of the cosmic SFRD obtained by \citet{madau2014}. The green dashed line, cyan dot-dashed line, and orange dotted line show the total (i.e., UV + IR) SFRD, UV SFRD, and IR SFRD obtained by \citet{burgarella2013}. Blue and brown squares are dust-uncorrected and -corrected SFRD obtained by \citet{bouwens2015}. {Blue open circles are results of \citet{rowan2016}. {Purple triangles and gray open circles indicate the cosmic SFRD obtained by the SCUBA2 large survey by \citet{cowie2017} and \citet{bourne2017}. Gray inverse-triangle are the contribution of bright ALESS sources \citep{swinbank2014}.} Brown open diamonds indicate the contribution of the ALMA sources obtained by \citet{dunlop2017}.} We note that these results are converted to the Chabrier IMF. 
\label{fig:sfrd}}
\end{figure*}

We estimate the contribution of the ASAGAO sources with the $K$-band counterparts to the cosmic SFRD. In Figure \ref{fig:sfrd}, we plot their contribution as a function of redshift. Here, we simply sum up the SFRs of the ASAGAO sources with the $K$-band counterparts and divide them by the co-moving volumes. {When we consider survey completeness obtained by \citet{hatsukade2018}, the contributions of ASAGAO sources to the cosmic SFRD are estimated to be $\sim$ 3 $\times$ 10$^{-2}$ and $\sim$ 8 $\times$ 10$^{-2}$ $M_\odot$ yr$^{-1}$ Mpc$^{-3}$ at 1 $<z<$ 2 and 2 $<z<$ 3, respectively. These values are consistent with results by \citet{hatsukade2018}.}

As a comparison, we plot the recent parametric descriptions of the redshift evolution of the cosmic SFRD obtained by \citet{madau2014}. Their results are based on the previous observations at UV-to-IR wavelengths. We also show the evolution of the cosmic SFRD at $z$ = 0--3.5 derived by \citet{burgarella2013}. They use UV and IR luminosity functions estimated by VIMOS-VLT Deep Survey (VVDS; \cite{lefevre2005}), {\it Herschel} large programs PACS evolutionary probe (PEP; \cite{lutz2011}), and {\it Herschel} Multi-tiered Extragalactic Survey (HerMES; \cite{oliver2012}) to estimate the cosmic SFRD. At $z=$ 3.8 and 4.9, we plot results of \citet{bouwens2015}. They assume UV luminosity functions estimated by {\it HST} data and dust correction based on the IRX-$\beta_\mathrm{UV}$ relation of \citet{meuer1999}. We also plot results of \citet{rowan2016}, which are based on {\it Herschel}-SPIRE 500 $\mu$m sources. {We also show the results of the SCUBA2 large survey by \citet{cowie2017} and \citet{bourne2017}, the contribution of bright ALESS sources by \citet{swinbank2014}, and the results from ALMA continuum surveys estimated by \citet{dunlop2017}. Figure \ref{fig:sfrd} suggests that the contribution of ASAGAO sources to the cosmic SFRD is $\gtrsim$ 1-dex larger than that of bright ALESS sources \citep{swinbank2014} at $z\sim$ 1--3.}

As shown in Figure \ref{fig:sfrd}, the shape of the contribution of the cosmic SFRD from the ASAGAO sources is similar to that of the previous observations. Figure \ref{fig:sfrd} shows that our results are $\gtrsim$ 1-dex smaller than the cosmic IR SFRD obtained by \citet{burgarella2013} at 1 $<z<$ 2. At these redshifts, galaxies with $\log(L_\mathrm{IR}/L_\odot)\lesssim$ 9--10 (i.e., fainter population than our ALMA detection limit) seems to be the main contributors to the cosmic IR SFRD. On the other hand, our results are consistent with the cosmic IR SFRD by \citet{burgarella2013} (Figure \ref{fig:sfrd}) at 2 $<z<$ 3. {This implies that in the redshift range of 2 $<z<$ 3, the most part of the cosmic IR SFRD 
predicted by \citet{burgarella2013} seems to be explained by ASAGAO sources.} {This can be a consequence of the evolution of the characteristic luminosities ($L^*_\mathrm{IR}$) of IR luminosity functions (i.e., at high redshift, $L^*_\mathrm{IR}$ becomes higher; \cite{hatsukade2018}).}
Nevertheless, the deduced IR SFRD by ASAGAO sources may be suffered from small statistics and/or field-to-field variance. A surveys with much wider survey volume would be necessary to mitigate such issues.

\section{Conclusions} \label{sec:summary}

In this paper, we report results of multi-wavelength analysis of ALMA 1.2-mm {detected ZFOURGE sources using ASAGAO data}. {We find that {24} ZFOURGE sources are detected by ALMA with S/N $>$ 4.5.} Their median redshift {($z_\mathrm{median}$ = 2.38 $\pm$ 0.14)} is consistent with redshifts of faint SMGs with $S_\mathrm{obs}\lesssim$ 1.0 mJy, although this value is lower than that of ``classical'' SMGs ($z_\mathrm{median}\sim$ 3.0). This difference can be caused by the redshift evolution of the IR luminosity function{, although we have to note that this can be caused by selection effect}.

Our SED fitting from optical to millimeter wavelengths suggest that ASAGAO sources mainly lie on the high-mass end of the main sequence of star-forming galaxies, although some ASAGAO sources show starburst-like features. On the other hand, the IRX-$M_*$, IRX-SFR, and IRX-$\beta_\mathrm{UV}$ relations of ASAGAO sources may imply that ALMA detected sources and non-detected sources have different dust properties (e.g., dust compositions or dust distribution) even if they show similar properties on the $M_*$--SFR plane. 

We resolve the contribution of dusty star-forming sources to the cosmic SFRD down to $\log(L_\mathrm{IR}/L_\odot)\sim$ 11, because of the high sensitivity and angular resolution of ALMA. We find that the ASAGAO sources with $K$-band counterparts are main contributors to the cosmic IR SFRD at 2 $<z<$ 3.

\begin{ack}
%
This paper makes use of the following ALMA data: ADS/JAO.ALMA\#2015.1.00098.S, 2015.1.00543.S, and 2012.1.00173.S.
ALMA is a partnership of ESO (representing its member states), NSF (USA), and NINS (Japan) together with NRC (Canada), NSC and ASIAA (Taiwan), and KASI (Republic of Korea) in cooperation with the Republic of Chile.
The Joint ALMA Observatory is operated by ESO, AUI/NRAO, and NAOJ.
Data analysis was partly carried out on the common-use data analysis computer system at the Astronomy Data Center (ADC) of the National Astronomical Observatory of Japan.
Y.~Yamaguchi is thankful for the JSPS fellowship.
This study was supported by the JSPS Grant-in-Aid for Scientific Research (S) JP17H06130 and the NAOJ ALMA Scientific Research Number 2017-06B.
H.~Umehata acknowledges support from the JSPS KAKENHI Grant Number JP20H01953.
Y.~Ao acknowledges support by NSFC grant 11933011.
\end{ack}

\appendix 
\section{The correspondence of IDs in previous papers to ASAGAO IDs} \label{sec:correspond}
\citet{ueda2018} and \citet{fujimoto2018} also report results of ASAGAO continuum sources. \citet{aravena2016}, \citet{dunlop2017}, \citet{franco2018}, and \citet{cowie2018} also observed the similar region of ASAGAO. In this section, we present the correspondence of their IDs to our ASAGAO IDs. There are no ALESS sources within the ASAGAO field.

\begin{table}[b!]
\footnotesize{
\caption{The correspondence to previous ALMA surveys in GOODS-S \label{tab:ID_goods-s}}
\begin{center}
\begin{tabular}{cc}
\hline\hline
{ASAGAO ID} & {ID in previous studies} \\
{(1)} & {(2)}\\
\hline
1 & UDF1, AGS6, SGS22, U3, F3\\
2 & AGS1, SGS5, U1, F1 \\
3 & AGS3, SGS9, U2, F2 \\
4 & UDF2, AGS18, SGS25, U6, F6 \\
5 & UDF3, ASPECS/C1, AGS12, SGS48, U8, F8 \\
6 & SGS20, U4, F4  \\
7 & SGS29, U5, F6  \\
8 & AGS13, SGS40, U12, F10 \\
9 & F9 \\
10 & UDF4, F132 \\
11 & F7 \\
12 & UDF5, F322 \\
13 & UDF6, F26 \\
14 & UDF7, U7 \\
15 & UDF11, F73 \\
16 & UDF8, ASPECS/C2, F90 \\
17 & U11 \\
19 & U10, F11 \\
23 & UDF13 \\
26 & SGS54, F103 \\
29 & F148 \\
31 & F113 \\
33 & F30 \\
44 & SGS63, F66 \\
\hline
\end{tabular}
\end{center}
\begin{tabnote}
{(1) ASAGAO IDs (2) Source IDs of ASPECS \citep{aravena2016}, UDF \citep{dunlop2017}, AGS \citep{franco2018}, SGS \citep{cowie2018}, U \citep{ueda2018}, and F \citep{fujimoto2018}.}
\end{tabnote}
}
\end{table}
\section{ASAGAO ID21}
\begin{figure*}[t!]
\begin{center}
\includegraphics[width=160mm]{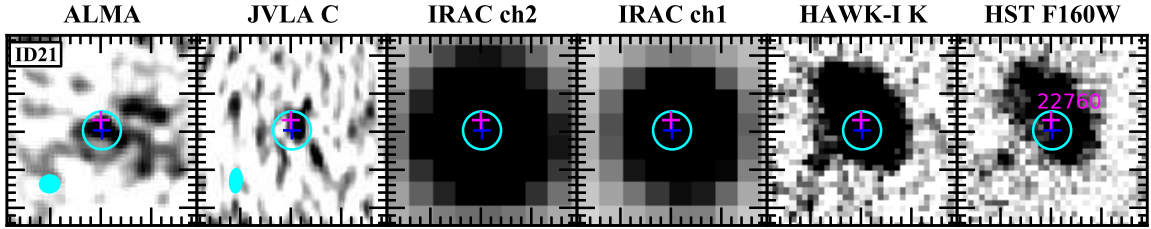}
\end{center}
\caption{Multi-wavelength images of ASAGAO ID21. From left to right: ALMA 1.2 mm, JVLA 6 GHz, \textit{Spitzer} IRAC/4.5 $\mu$m, IRAC/3.6 $\mu$m, VLT HAWK-I/$K_s$, and HST WFC3/$F160W$ images. The field of view is 5$^{\prime\prime}$ $\times$ 5$^{\prime\prime}$. Blue and magenta crosses mark the ALMA ositions and ZFOURGE positions, respectively. Cyan circles are 1$^{\prime\prime}$ apertures. The synthesized beams of ALMA and JVLA are expressed as cyan ellipses.
\label{fig:ID21_postagestamp}}
\end{figure*}

\begin{figure}[t!]
\begin{center}
\includegraphics[width=80mm]{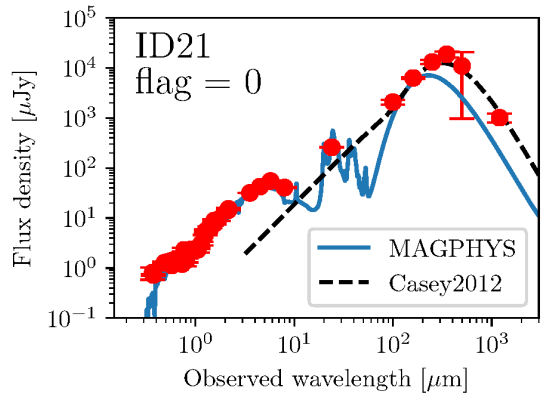}
\end{center}
\caption{The estimated SEDs of ASAGAO ID21. Red symbols with errors are observed flux densities. Blue solid lines are best fit SEDs estimated by \texttt{MAGPHYS}. The black dashed lines are the mid-IR to far-IR SED model by \citet{casey2012}.
\label{fig:ID21_SED}}
\end{figure}

We show the multi-wavelength postage stamp of ASAGAO ID21 in Figure \ref{fig:ID21_postagestamp}. ASAGAO ID 21 has ``\texttt{use flag} $=$ 0'' in \citet{straatman2016}. This is the reason why we remove this source from our analysis. However, we note that it has a spectroscopic redshift obtained by \citet{wisnioski2015} ($z_\mathrm{spec}$ = 2.187). In the case that we adopt this redshift and run the \texttt{MAGPHYS}, the best fitted SED is shown in Figure \ref{fig:ID21_SED}. Figure \ref{fig:ID21_SED} shows that the stellar light is dominating the fit and far-IR to millimeter spectrum is hugely underpredicted. This implies that the far-IR to millimeter bright region can lie at higher redshift than the optical/near-IR identified region. Chance alignment of (and perhaps associated gravitational amplification of) a dusty background galaxy with a physically unrelated galaxy in the foreground (e.g., \cite{bourne2014,oteo2017}) could be responsible for the catastrophic SED fit.

\end{document}